\documentclass[pre,twocolumn,showpacs,floatfix]{revtex4}
\usepackage{graphicx}
\usepackage[usenames]{color}
\newcommand{\beq}{\begin{equation}}
\newcommand{\eeq}{\end{equation}}
\newcommand{\beqa}{\begin{eqnarray}}
\newcommand{\eeqa}{\end{eqnarray}}
\newcommand{\ba}{\begin{array}}
\newcommand{\ea}{\end{array}} 
\begin{document} 

\title{Superfluid Fermi-Fermi mixture:
phase diagram, stability, and  soliton formation} 
\author{Sadhan K. Adhikari
\footnote{Email: adhikari@ift.unesp.br; 
URL: www.ift.unesp.br/users/adhikari}} 
\affiliation{Instituto de F\'isica Te\'orica, UNESP $-$ S\~ao Paulo
State University,
01.405-900 Sao Paulo, Sao Paulo, Brazil }

\begin{abstract} We study the phase diagram for a dilute
Bardeen-Cooper-Schrieffer superfluid Fermi-Fermi mixture (of distinct
mass) at zero temperature using energy densities for the superfluid
fermions in one (1D), two (2D), and three (3D) dimensions.  We also
derive the dynamical time-dependent nonlinear Euler-Lagrange equation
satisfied by the mixture in one dimension using this energy density.  We
obtain the linear stability conditions for the mixture in terms of
fermion densities of the components and the interspecies Fermi-Fermi
interaction. In equilibrium there are two possibilities. The first is
that of a uniform mixture of the two components, the second is that of
two pure phases of two components without any overlap between them.  In
addition, a mixed and a pure phase, impossible in 1D and 2D, can be
created in 3D. We also obtain the conditions under which the uniform
mixture is stable from an energetic consideration.  The same conditions
are obtained from a modulational instability analysis of the dynamical
equations in 1D.  Finally, the 1D dynamical equations for the system are
solved numerically and by variational approximation (VA) to study the
bright solitons of the system for attractive interspecies Fermi-Fermi
interaction in 1D. The VA is found to yield good agreement to the
numerical result for the density profile and chemical potential of the
bright solitons.  The bright solitons are demonstrated to be dynamically
stable. The experimental realization of these Fermi-Fermi bright
solitons seems possible with present setups.  \end{abstract}

\pacs{ 03.75.Ss,  64.75.+g, 03.75.Kk}

\maketitle

\section{Introduction}

After the experimental realization of a trapped Bose-Einstein condensate 
(BEC) \cite{books} 
there has been a great effort to trap and cool the Fermi atoms to 
degeneracy by sympathetic cooling in the presence of a second Bose or 
Fermi component. The second component is needed to facilitate 
evaporative cooling not possible due to lack of interaction in a 
single-component Fermi gas \cite{exp1,exp2}. 
Apart from the observation of  the
degenerate Bose-Fermi
mixtures 
 $^{6,7}$Li
\cite{exp3,mix-ex}, $^{23}$Na-$^6$Li \cite{exp4} and $^{87}$Rb-$^{40}$K
\cite{exp5,exp5x}, there have been studies of the following
spin-polarized
degenerate Fermi-Fermi mixtures 
 $^{40}$K-$^{40}$K \cite{exp1} and $^6$Li-$^6$Li \cite{exp2} in 
different hyperfine states.

Specially challenging has been the 
experimental realization of vortex lattice in a 
Bardeen-Cooper-Schrieffer (BCS)
superfluid Fermi gas 
\cite{exp6,exp7,bcsexp,fw} in Bose-Fermi mixture 
employing   a weak 
attractive interaction among the intra-species fermions by using a 
Feshbach resonance \cite{fesh,FESH}. This attractive  interaction allows 
the 
formation of 
BCS pairs leading to a BCS superfluid \cite{fw,leggett-book}. 
In the last few years 
by further increasing this attraction 
several
experimental groups have observed 
the crossover \cite{cross}
from the paired 
BCS 
state  to the BEC 
of molecular dimers with ultra-cold two-hyperfine-component Fermi
vapors of $^{40}$K  \cite{greiner}
and $^6$Li atoms \cite{zwierlein,chin}. 
Another possibility is to use two distinct Fermi atoms for this purpose
as suggested in Ref. \cite{skac} in a study of collapse in a Fermi-Fermi 
mixture
($^6$Li-$^{40}$K is a possible candidate for future exploration.) 
The Feshbach-resonance management of the Fermi interaction  could be 
utilized to study a superfluid Fermi-Fermi 
mixture in a controlled fashion \cite{fesh}.

In Bose-Fermi mixtures, there have been several studies on 
 phase 
separation 
\cite{molmer,pethick,viverit,das,sala-toigo}, 
 soliton-like structures \cite{Sadhan-BFsoliton}, and collapse 
\cite{skac2},   
recently.
The phase diagram of the Bose-Fermi mixture in three dimensions (3D) has 
been studied by 
Viverit {\it et al.} \cite{viverit}, whereas the same in one dimension 
(1D) has been 
studied by Das \cite{das}.   
Bright solitons have been observed in BECs of Li \cite{Li-soliton} and 
Rb \cite{Rb-soliton} atoms and studied subsequently 
\cite{BECsolitons}. It has been demonstrated using  microscopic 
\cite{BFsoliton(BBrepBFattr)} and mean-field hydrodynamic 
\cite{Sadhan-BFsoliton} models 
that the formation of stable fermionic bright solitons is 
possible in a
degenerate Bose-Fermi  mixture
in the presence of a sufficiently
attractive interspecies interaction which can overcome the 
Pauli-blocking repulsion
among fermions. The formation of a soliton in these cases is
related to the fact that the system can lower its energy by forming high 
density region (bright soliton) 
 when the interspecies attraction is large
enough to overcome the Pauli-blocking interaction in the degenerate 
Fermi gas  (and 
any possible repulsion in the BEC) \cite{skbs}. 
There have also been studies of mixing-demixing transition 
in degenerate Bose-Fermi \cite{sala-sadhan2}
and Fermi-Fermi \cite{skabm} mixtures, and soliton formation in 
Fermi-Fermi mixtures \cite{ff} 
.

In this paper we investigate the phase diagram of a BCS superfluid
Fermi-Fermi mixture of fermion components  of distinct mass
at zero temperature  
using energy densities for the
superfluid Fermi components in one, two (2D), and three  dimensions. 
We derive the conditions
of stability of the mixture in terms of the densities of the components
and the strength of interspecies interaction. The two possible phases of
the mixture are a uniformly mixed configuration and a totally separated
pure-phase configuration. Unlike in a
Bose-Fermi mixture \cite{viverit,das},
no
complicated mixed phases
are allowed in a superfluid Fermi-Fermi mixture in 1D and 2D. However, a 
mixed and a pure phase is allowed in 3D. 
In 1D, two 
pure and 
separated phases of the fermion 
components appear for low fermion densities, whereas the opposite is 
found in 3D. In 1D, a uniform mixture appears  for large fermion 
densities with the opposite taking place in 3D. In 2D, the condition for 
uniform mixture and phase separation   is independent of density of the 
components. 
In 1D, we find the uniform mixture to be unstable for small fermion 
densities, whereas  In 3D, the uniform mixture is unstable for large 
fermion
densities.

The 1D configuration is of special interest due to soliton
formation by modulational instability 
of a uniform mixture.  To study
this phenomenon we derive a set of dynamical equations of the system as 
the 
Euler-Lagrange equation of an appropriate Lagrangian.  The
condition of stability of the uniform mixture and the formation of 
soliton for attractive interspecies Fermi-Fermi interaction 
were studied from an energetic consideration as well as  with a linear
stability analysis of the constant-amplitude solution of the above 
dynamical
equations. We solved the 1D dynamical equations numerically and
variationally to study some features of the bright solitons. The 
numerical
results for  the density of the fermion components as well as their
chemical potentials are found to be in good agreement with the
variational findings. These bright solitons are found to be stable 
numerically when they are subjected to a
perturbation.

The dependence of 
Fermi energy densities in 1D and 2D on atomic densities 
has counterparts in Bose systems and the analysis presented here is also 
applicable to these Bose systems. The 2D Fermi energy density has a 
quadratic dependence on atomic density as in a dilute BEC obeying the 
Gross-Pitaevskii equation, thus allowing the present results to be 
applicable to such 
a BEC \cite{books}. The 1D Fermi energy density, on the other hand, has 
a 
cubic dependence on atomic density as in a Tonks-Girardeau \cite{Tonks} 
(TG) 
Bose 
gas 
observed recently \cite{tg}, thus making the present results applicable 
to 
this 
system.

The paper is organized as follows. 
In Sec. II we consider the stability condition of a uniform BCS 
superfluid
Fermi-Fermi mixture from an energetic consideration.  
In Sec. III we consider a two-phase BCS superfluid Fermi-Fermi mixture
in 1D, 2D and 3D and study the possibility of the formation of two 
phases 
from a consideration of pressure, energy and chemical 
potential of the system. We can have  two pure phases or a uniformly 
mixed 
phase in all dimensions.  In addition, in 3D, we can have a pure and a 
mixed phase. 
In Sec. IV we consider the Euler-Lagrange nonlinear dynamical equations 
for the system in 1D and study the modulational instability of the 
constant-amplitude solution representing the uniform mixture. The 
condition of modulational instability for attractive Fermi-Fermi 
interaction 
is found to be consistent with the condition of stability of the uniform 
mixture obtained from an energetic consideration in Sec. II. We further 
solve these dynamical equations numerically and variationally to analyze
the properties of the Fermi-Fermi solitons.   
Finally, in Sec. V we present a summary of our study. 

\section{Uniform Superfluid Fermi-Fermi Mixture}

\subsection{Energy Density of a Component}

We consider a single-component dilute BCS superfluid of spin-half  Fermi 
atoms 
of  mass $m$ and density $n_3$
with a weak attraction
between fermions with opposite spin  orientations. 
In 3D, the energy density of this system is given by  
\cite{yang1,yang2,heis,salasnich},
\begin{equation}
\mathcal{E}_{\mathrm{3D}}=(3/5){n}_{\mathrm{3}}\varepsilon _{F}
\label{1A}
\end{equation}%
where $\varepsilon _{F}=\left( \hbar k_{F}\right) ^{2}/(2m)$ is the 
Fermi
energy, $\hbar k_{F}$ is the Fermi momentum  
(this expression was first obtained by Lee and Yang 
\cite{yang2} in the weak-coupling BCS limit). Modifications to this 
expression
for a description of the BCS-BEC crossover, for stronger attraction 
between
fermions, have also been considered \cite{salasnich}. 
The total density of the fermions in a 3D box is obtained by filling the 
quantum states up to the Fermi energy and  
is
given by ${n}_{\mathrm{3}}={2(2\pi )^{-3}}\int_{0}^{k_{F}}4\pi
k^{2}dk\equiv \left( 3\pi ^{2}\right) ^{-1}\left( 2m\varepsilon 
_{F}/\hbar
^{2}\right) ^{3/2}$. (The factor of 2 in the expression for $n_3$ 
accounts for BCS pairing in each level.)
Hence the energy density in  (\ref{1A}) becomes
\begin{equation}
\mathcal{E}_{\mathrm{3D}}=\frac{3(3\pi ^{2})^{2/3}\hbar 
^{2}}{10m}{n}%
_{3}^{5/3} = \frac{3}{5}A_{\mathrm{3}}{n}_{3}^{5/3} ,  \label{1C}
\end{equation}
with $A_ {\mathrm{3}}=\hbar^2(3\pi^2)^{2/3}/(2m)$.

Similarly, the energy density of a dilute  1D superfluid of atom density 
$n_1$ 
is given by
\cite{yang3,recati}
\begin{equation}
\mathcal{E}_{\mathrm{1D}}=(1/3)n_{\mathrm{1}}\varepsilon 
_{F}.
\label{1E}
\end{equation}%
This was obtained using the Gaudin-Yang (GY) model \cite{yang3} of 
fermions 
weakly
interacting via zero-range ($\delta$-function) potential, and was later 
extended to
the description of the BCS-to-unitarity crossover  \cite{recati}. 
({For repulsive interaction the GY model gives \cite{xt} 
a 
Tomonaga-Luttinger 
liquid \cite{TL}, while for attractive interaction it leads to a 
Luther-Emery liquid \cite{LE}. For weak attraction the ground 
state of the system is a BCS superfluid \cite{fw,KO}. With the increase 
of 
attraction,  the strong-coupling 
regime 
of tightly bound dimers is attained, which behaves like a hard 
core Bose gas, or like a 1D noninteracting Fermi gas, 
known as the TG gas \cite{Tonks,ad-sa}.) The general solution for the 
ground-state energy in the GY model has been  obtained by solving the 
Bethe 
ansatz \cite{bethe} equations  for all strengths of $\delta$ 
interaction connecting the weak-attraction regime of BCS condensate to
 the strong-attraction regime of of tightly bound dimers described by 
the Lieb-Liniger model \cite{ll}  of repulsive bosons. This solution can 
be presented as an expansion series in limits of weak or strong 
interactions. 
The limiting value of this solution in the weak interaction 
BCS limit is given by Eq. (\ref{1E}) \cite{recati,xt}.}

The fermion density 
of
the BCS superfluid in a 1D box  is 
$n_{\mathrm{1}%
}={2(2\pi )}^{-1}\int_{-k_{F}}^{+k_{F}}dk\equiv \left( 2/\pi \hbar 
\right)
\sqrt{2m\varepsilon _{F}}$, hence, in this case, $\varepsilon _{F}=\pi
^{2}\hbar ^{2}n_{1}^{2}/(8m)$, and energy density (\ref{1E})
becomes \cite{recati2}
\begin{equation}
\mathcal{E}_{\mathrm{1D}}=\frac{\pi ^{2}\hbar 
^{2}}{24m}n_{\mathrm{%
1}}^{3}=\frac{1}{3}A_\mathrm{1} n_\mathrm{1}^3,
  \label{1G}
\end{equation}
with $A_\mathrm{1}=\hbar^2\pi^2/(8m)$. The energy density of a 
TG gas \cite{Tonks} is given by $\mathcal{E}_{\mathrm{TG}}
= \hbar^2 \pi^2 n_1^3/6m$ \cite{ad-sa} and is very similar to that given 
by Eq. 
(\ref{1G}). The difference in numerical factors between the two 
expressions is due to pairing in the present Fermi superfluid allowing 
two fermions (spin up and down) in the same quantum level. Hence the 1D 
results of the present study is also applicable to a TG 
gas. 

Finally, a counterpart of relations (\ref{1A}) and (\ref{1G}) for the 2D
superfluid is \cite{luca2}
$\mathcal{E}_{\mathrm{2D}}=(1/2)n_{\mathrm{2}%
}\varepsilon _{F}$, the 2D density being 
$n_{\mathrm{2}}={2(2\pi
)^{-2}}\int_{0}^{k_{F}}2\pi kdk\equiv \left( m/\pi \hbar ^{2}\right)
\varepsilon _{F}$, with  $\varepsilon _{F}=\pi \hbar ^{2}n_{%
\mathrm{2}}/m$. Thus, the energy density of the 2D superfluid can 
be written  as \cite{luca2} 
\begin{equation}
\mathcal{E}_{\mathrm{2D}}=\frac{\pi \hbar 
^{2}}{2m}n_{\mathrm{2}%
}^{2} = \frac{1}{2} A_\mathrm{2}n_{\mathrm{2}}^2,  \label{1J}
\end{equation}
with $A_{\mathrm{2}}=\pi\hbar^2/m$.

{Here we specify the criteria of applicability of 
Eqs. 
(\ref{1C}), (\ref{1G}), and (\ref{1J}) for different dimensionalities. 
These results are valid for a dilute BCS superfluid. In 3D, at low 
densities, $k_F|a_F|<<1$ with $a_F$ the
Fermi-Fermi scattering length, 
gaps 
are small and have little effect on the total energy of the system 
\cite{heis}. The 
total energy density of the ground state can then be expanded in powers 
of the small parameter $k_F|a_F|$. At low densities Eq. 
(\ref{1C}) includes the lowest order term in this expansion 
\cite{yang2}. The condition  $k_F|a_F|<<1$ of validity of Eq. 
(\ref{1C})
can be related to the gas 
parameter $n_3|a_F|^3$ in 3D: $n_3|a_F|^3<<1/(3\pi^2)$, as the density 
$n_3 
= 
k_F^3/(3\pi^2)$.  In 1D, for a $\delta$ interaction of strength $g_1$ 
the dimensionless coupling constant $\gamma=mg_1/(\hbar^2 n_1)$ and the 
condition of validity of Eq.
(\ref{1G}) is $|\gamma|<<1$ \cite{recati}.  
In two dimensions an attractive interaction leads to a bound state of 
energy $\epsilon_0$ and the condition of diluteness for the validity of 
Eq. (\ref{1J}) can be expressed as $\epsilon_0/\varepsilon_F 
<<1.$}

\subsection{Stability Condition of the Uniform  
Mixture}

We consider a uniform mixture 
of two types of fermions, containing $N_i, 
i=1,2$,  atoms (of mass $m_1=m$ and $m_2=m/\lambda$) ,  in a box of 
size $S$  (in 1D the size is a length, in 2D an area, and in 3D a 
volume)
with distinct mass at zero 
temperature. 
The energy density of the uniform mixture 
is given by 
\begin{eqnarray}\label{en-den1}
{\cal E}_{\mathrm{1D}} &=& \frac{1}{3}A_\mathrm{1} n_{1(1)}^3+ g_{12} \, 
n_1 
\, 
n_2+\frac{1}{3}\lambda A_\mathrm{1} n_{1(2)}^3,\\
\label{en-den2}
{\cal E}_{\mathrm{2D}} &=& \frac{1}{2}A_\mathrm{2} n_{2(1)}^2+ g_{12} \, 
n_1 
\,
n_2+\frac{1}{2}\lambda A_\mathrm{2} n_{2(2)}^2,\\
{\cal E}_{\mathrm{3D}} &=& \frac{3}{5}A_\mathrm{3} n_{3(1)}^{5/3}+ 
g_{12} \, 
n_1 \,
n_2+\frac{3}{5}\lambda A_\mathrm{3} n_{3(2)}^{5/3},
\label{en-den3}
\end{eqnarray}
respectively, for 1D, 2D and 3D systems, where $n_{d(i)}=N_i/S$ denotes 
the 
density of each component in $d$D, $d=1,2,3$. 
The nonlinear terms involving $g_{12}=4\pi\hbar^2a_{12}/m_{12}$ in above 
equations  represent the interaction between two types of atoms 
arising solely from the atomic scattering length
 $a_{12}$, where $m_{12}$ is the reduced mass of atoms.
The terms involving $A_j$ in the above equations, although is similar to  
the  $gn^2/2$ interaction term for bosons (with $g=4\pi\hbar^2 
a/m$ representing the self interaction of a dilute boson gas with $a$ 
the Bose-Bose scattering length and $m$ the mass  of an atom), have a 
different origin as we have seen. These terms  originating  from the 
energy of the 
fermions occupying the lowest quantum levels at zero temperature obeying 
Pauli principle generate an effective repulsion between the 
fermions and 
is usually called Pauli-blocking interaction.

The chemical potentials $\mu_i\equiv \partial {\cal E}/\partial n_i$ 
for species  $i=1,2$ in 1D, 2D and 3D, are given, respectively,  by
\begin{eqnarray}\label{mu1D}
\mu_1&=&A_\mathrm{1}n_{1(1)}^2+g_{12}n_{1(2)}, \quad \mu _2= 
g_{12}n_{1(1)}+\lambda  
A_\mathrm{1}n_{1(2)}^2,\nonumber  \\ \\
\label{mu2D}
\mu_1&=&A_\mathrm{2}n_{2(1)}+g_{12}n_{2(2)}, \quad \mu _2= 
g_{12}n_{2(1)}+\lambda  
A_\mathrm{2}n_{2(2)},\nonumber \\ \\
\label{mu3D}
\mu_1&=&A_\mathrm{3}n_{3(1)}^{2/3}+g_{12}n_{3(2)}, \quad \mu _2= 
g_{12}n_{3(1)}+\lambda  
A_\mathrm{3}n_{3(2)}^{2/3}.\nonumber \\
 \end{eqnarray}

The uniformly  mixed phase is energetically stable if its energy is a 
minimum 
with respect to small variations of the densities, while the 
total number of fermions and bosons are held fixed. The conditions of 
stability [are the conditions of a minimum of ${\cal 
E}(n_{(1)},n_{(2)})$ as a 
function of two variables $n_{(1)}$ and $n_{(2)}$ and]  
are given by 
\begin{eqnarray}\label{A1}
&\frac{\partial^2 {\cal E}}{\partial n_{(1)}^2}\equiv 
\frac{\partial \mu_1}{\partial n_{(1)}}\ge 0, \quad 
\frac{\partial^2 {\cal E}}{\partial n_{(2)}^2}\equiv
\frac{\partial \mu_2}{\partial n_{(2)}}\ge 0,&  \\
\label{A2}
&\frac{\partial^2 {\cal E}}{\partial n_{(1)}^2}
\frac{\partial^2 {\cal E}}{\partial n_{(2)}^2}- \left( \frac{\partial^2 
{\cal E}}{\partial n_{(1)}
\partial n_{(2)}}
   \right)^2 \equiv
\frac{\partial \mu_1}{\partial n_{(1)}}\frac{\partial \mu_2}{\partial 
n_{(2)}}-\frac{\partial \mu_1}{\partial n_{(2)}}\frac{\partial 
\mu_2}{\partial 
n_{(1)}}
\ge 0,& \nonumber \\ 
\end{eqnarray}
where we have dropped the  dimension suffix.
The solution of these inequalities gives the region 
in the parameters' space where the uniformly  mixed
phase is energetically  stable. Using Eqs. (\ref{A1}) and 
(\ref{A2})
the condition of stability of the 
uniform mixture in 1D, 2D, and 3D are given, respectively, by
\cite{viverit,ad-sa}
\begin{eqnarray}\label{st1}
&&4A_\mathrm{1}^2\lambda n_{(1)}n_{(2)} \ge g_{12}^2, \\
\label{st2}
&&A_\mathrm{2}^2\lambda \ge g_{12}^2 , \\
\label{st3}
&&4A_\mathrm{3}^2 \lambda \ge 9 g_{12}^2n_{(1)}^{1/3}n_{(2)}^{1/3}. 
\end{eqnarray}
These conditions are determined by $ g_{12}^2$ and not the sign of 
$g_{12}$. 

In 1D, we find from Eq. (\ref{st1})
with a finite $g_{12}^2$, that at small fermionic densities (small  
$n_{(1)}$ and $n_{(2)}$) 
the uniform mixture is unstable: 
the ground-state of the system displays demixing if $g_{12}>0$
and becomes a localized Fermi-Fermi bright soliton if $g_{12}<0$
\cite{ad-sa}.
The mixture is stable at large fermionic densities. 
In 2D, Eq. (\ref{st2}) reveals that the condition for stability is 
independent of density. In 3D, Eq. (\ref{st3}) predicts that 
for a finite $g_{12}^2$, the mixture is unstable at large 
fermionic densities, leading to collapse for $g_{12}<0$ and to demixing 
for $g_{12}>0$,  and stable at small fermionic densities.
It is realized that as we move from 1D to 3D through 2D, 
the condition of stability  of the uniform mixture 
 changes from large fermion densities to small fermion densities.  
This result is quite similar to that in a Bose-Fermi 
mixture \cite{viverit,das}, where the 
condition of stability of the uniform mixture is  independent of the 
bosonic density and has a similar dependence on fermion density, e.g., 
during the passage from 1D to 3D through 2D,
the condition of stability changes from large fermion density to small 
fermion density.

 From inequality (\ref{A1})
the stability condition of a single component uniform gas can be 
represented as $\partial \mu_1 /\partial n_1 >0$, which, using Eqs. 
(\ref{mu1D}), 
(\ref{mu2D}) 
and (\ref{mu3D}),
 is 
realized for $A_d>0$ denoting a repulsive system. 
In the presence of a second component, 
inequality (\ref{A2}) can be written as \cite{viverit}
\begin{eqnarray}\label{A22}
\frac{\partial \mu_1 }{\partial n_{(1)}}-\left(\frac{\partial 
\mu_2}{\partial n_{(1)}}\right)^2\frac{\partial n_{(2)}}{\partial \mu_2} 
\ge 0,
\end{eqnarray}
as ${\partial \mu_2 }/{\partial n_{(1)}}={\partial \mu_1 }/
{\partial 
n_{(2)}}$.
The first term ${\partial \mu_1 }/{\partial n_{(1)}}$
in inequality (\ref{A22}) represents the effective 
repulsion among fermions of type 1. 
The second term, representing an induced interaction due to the presence 
of component 2,    
reduces  the 
repulsion and tries to destabilize the uniform mixture. The uniform 
mixture becomes unstable when the second term in inequality (\ref{A22}) 
becomes larger than the first term. This 
happens for both attractive and repulsive interspecies interaction 
$g_{12}={\partial \mu_2 }/{\partial n_{(1)}}$.

The inequality (\ref{A2}) can be written as 
\beq
c_1^2 c_2^2 \ge 4 g_{12}^2 n_{(1)} n_{(2)} ,  
\label{ineq2} 
\eeq
where 
$c_i=\sqrt{2 n_{(i)} (\partial \mu_i/\partial n_{(i)})}$
represent sound velocities in the two superfluid components, $i=1,2$. 
The sound velocity $c_{12}$ of the 1D Fermi-Fermi mixture can be 
obtained following a procedure suggested 
by Alexandrov and Kabanov \cite{kabanov,ad-sa} 
for a two-component BEC:
\beq
c_{12}= {\frac{1}{\sqrt{2}}} \sqrt{c_1^2+c_2^2 \pm 
\sqrt{(c_1^2-c_2^2)^2 + 16 g_{12}^2 n_{(1)} n_{(2)} }} \; . 
\eeq
The homogeneous mixture 
becomes  unstable when the sound velocity $c_{12}$ becomes imaginary, 
e.g., when inequality (\ref{ineq2}) is violated.

\section{Two-Phase Superfluid Fermi-Fermi Mixture}

In the last section we considered a uniform mixture of two 
components in equilibrium. Here  we explore the more interesting 
case of two types of fermions with different possible densities in different regions of a box of size $S.$  The components 
may mix uniformly or form separate phases depending on the initial 
conditions $-$ mass, density, interspecies interaction etc.

The conservation of the number of particles, $N_1$ and $N_2$, of the two 
species can be 
expressed as \cite{viverit,das}
\beq
N_i=Sn_i={S}
\sum_{j=1}^2 n_{i,j}f_j, \quad \sum_{j=1}^2 f_j=1,
\eeq
where $i=1,2$ represent the species and $j=1,2$ represent the phases 
(different region with distinct density of gas),
$n_i=N_i/S$ represent the overall density of the two species,  
$ n_{i,j}$ is the density of species $i$ in phase $j$,
and  $S_j=Sf_j$ represent the size  of each phase with $f_j$ the 
fraction 
of size in phase $j$. For a two-component system 
one can have only two distinct phases, $j=1,2$,  as the inclusion of 
more phases leads to inconsistency \cite{viverit}. Here we have dropped 
the dimension label $d$ and also removed the parentheses () from the 
component label $i$.

The total energy of the system is given by   
\begin{eqnarray}\label{A23}
{E}&=&\sum_{j=1}^2 {E}_j \equiv  S\sum_{j=1}^2f_j{\cal E}_j
\end{eqnarray}
where ${\cal E}_j$ denotes the energy density of phase $j$ and ${ 
E}_j$ its total energy. The pressure $P_j$
of phase $j$ is given by $P_j=-\partial { E}_j/\partial S_j$. The chemical potential of component 
$i$ in phase $j$ is defined by $\mu_{i,j}=\partial {\cal E}_j/\partial 
n_{i,j}$. 

For equilibrium, the pressure in one phase has to be equal to that in 
the 
other. If two phases are occupied by atoms of the same type, the 
chemical potential for that type of atoms in two phases should also be 
equal so that the equilibrium can be energetically maintained. If the 
atom density of one type of atom in a phase is zero then the chemical 
potential of that type of atom in this phase should be larger than  
that in the other, so that the atoms do not flow to the phase with 
no atoms of this type \cite{viverit}.

In the following we consider a system composed of two phases comprising 
of fractions $f_1 =f$ and $f_2=(1-f)$ of size $S$. There are three 
following possibilities to be analyzed in 1D, 2D, and 3D, although some 
of 
them  
may not materialize in a particular case:

(i) Two pure and separated phases  with one
type of atom occupying a distinct phase.

(ii) A mixed and a pure phase where the density of one type of atoms is 
zero in one phase.

(iii) Two mixed phases where both phases are occupied by both type of 
atoms.

In the following we deal with the three possibilities in 1D, 2D and 3D. 
First, we consider the 2D case as the algebra is significantly simpler 
in this case.

\subsection{Two-Dimensional (2D) Mixture}

From Eqs. (\ref{en-den2}) we find that 
the expressions for total energy and pressure  in this case are 
\begin{eqnarray}\label{p1}
 { E}_j&=&S_j {\cal E}_j \equiv 
S_j\left[  
\frac{1}{2}A_\mathrm{2}(n_{1,j}^2+\lambda 
n_{2,j}^2)+g_{12}n_{1,j}n_{2,j} \right],\\
\label{p2}
P_j&\equiv & -\frac{\partial { E}_j}{\partial S_j}=
 \frac{1}{2}A_\mathrm{2}n_{1,j}^2+g_{12}n_{1,j}n_{2,j}+
 \frac{1}{2}A_\mathrm{2}\lambda n_{2,j}^2.
\end{eqnarray} 
In deriving Eq. (\ref{p2}) we recall that $n_{i,j}\sim 1/S_j$. From Eq. 
(\ref{p1}) the chemical potentials are given by
\begin{eqnarray}\label{p3}
\mu _{1,j}&=& A_\mathrm{2} n_{1,j} +g_{12}n_{2,j} \\
\label{p4}
\mu _{2,j}&=& A_\mathrm{2}\lambda  n_{2,j} +g_{12}n_{1,j} 
\end{eqnarray}

\subsubsection{Two Pure Phases}

In case of two pure and separated phases one should have, for 
example $n_{1,2}=n_{2,1}=0$ corresponding to the type one atoms 
occupying phase 1 only ($n_{1,1}\ne 0$)
and type 2 atoms occupying phase 2 only ($n_{2,2}\ne 0$).

Equality of pressure $P_1=P_2$ in the two phases yields
\beq \label{p5}
 n_{1,1}^2 = \lambda n_{2,2}^2.
\eeq
As the number of atoms is zero in one of the phases, one has the 
inequalities $\mu_{2,2}\le \mu_{2,1}$ and $\mu_{1,1}\le \mu_{1,2}$
on the  chemical potential, which, using Eqs. (\ref{p3}) and (\ref{p4}),  
become
\beqa \label{p6}
A_\mathrm{2}\lambda n_{2,2} &\le & g_{12}n_{1,1}, \\
A_\mathrm{2} n_{1,1} &\le & g_{12}n_{2,2}. 
\label{p7}
\eeqa
Eliminating $n_{1,1}$ and $n_{2,2}$ among Eqs. (\ref{p5}), (\ref{p6}), 
and  
(\ref{p7}) we get 
\beqa \label{p8}
A_\mathrm{2}^2\lambda \le g_{12}^2,
\eeqa
consistent with inequality (\ref{st2}). We have the uniform 
mixture 
for 
inequality (\ref{st2}); for the opposite inequality (\ref{p8})
we have the separated phases in equilibrium.  These 
inequalities  are  independent of the atomic densities. 

In the present case the overall densities of the two species are given 
by
\beqa  \label{q1}
n_1=f_1 n_{1,1} =f n_{1,1}, \quad n_2=f_2 n_{2,2} =(1-f)n_{2,2}.
\eeqa

Let us now consolidate these findings using energetic considerations 
comparing the total energy of a phase-separated configuration with that 
of a uniform mixture. The energy of the mixture is given by
\beqa
{ E}_{\mathrm{mix}}&=& S\left[\frac{1}{2}A_\mathrm{2} n_{1}^2  
+g_{12}n_1n_2
+\frac{1}{2}A_\mathrm{2} \lambda n_{2}^2 
  \right],\\
&=& S\biggr[  \frac{1}{2}A_\mathrm{2} f^2 n_{1,1}^2  
+g_{12}n_{1,1}n_{2,2}f(1-f) 
\nonumber \\
&+& \frac{1}{2}A_\mathrm{2} \lambda  n_{2,2}^2 (1-f)^2 
   \biggr],
\eeqa
where we have used Eqs. (\ref{q1}).
The energy of the phase-separated system with the same number of atoms 
is 
\beqa
{ E}_{\mathrm{sep}}&=& S \left[ \frac{1}{2}A_\mathrm{2} f   
n_{1,1}^2
+ \frac{1}{2}A_\mathrm{2} \lambda (1-f) n_{2,2}^2 \right]. 
\eeqa
Using Eq. (\ref{p5}), one has for the difference
\beqa
{ E}_{\mathrm{mix}}-
{ E}_{\mathrm{sep}}=Sf(1-f)n_{2,2}^2\sqrt \lambda 
[g_{12}-A_\mathrm{2}\sqrt 
\lambda ].
\eeqa
When ${ E}_{\mathrm{mix}}> { E}_{\mathrm{sep}}$ the system 
naturally moves to the separated phase and this happens for $g_{12}^2
>A_\mathrm{2}^2 \lambda $, consistent with inequality (\ref{p8}), 
leading to a 
stable separated phase. In the opposite limit, when ${ 
E}_{\mathrm{mix}}< { E}_{\mathrm{sep}}$, the energetic 
consideration favors the uniform mixture and this happens for 
$g_{12}^2<A_\mathrm{2}^2 \lambda $, consistent with inequality 
(\ref{st2}).

\subsubsection{A mixed and a pure phase}

Here  we consider  one mixed phase (phase 1) and one 
pure phase (phase 2)
consistent with $n_{1,2}=0$, which means that the type 1 atoms 
occupy 
only phase 1, whereas type 2 atoms occupy both phases 1 and 2.
Using Eq. (\ref{p2})
the equality of pressure in two phases leads to  
\beqa\label{q5}
\frac{1}{2}A_\mathrm{2}n_{1,1}^2+g_{12}n_{1,1}n_{2,1}+
\frac{1}{2}A_\mathrm{2}\lambda n_{2,1}^2= \frac{1}{2}
A_\mathrm{2}\lambda n_{2,2}^2.
\eeqa
From Eq. (\ref{p4})
the equality of the chemical potential of type 2 atoms in two phases 
($\mu _{2,2}=\mu_{2,1}$)
leads to 
\beqa\label{q6}
n_{1,1}=A_\mathrm{2}\lambda (n_{2,2}-n_{2,1})/g_{12} .
\eeqa
From Eq. (\ref{p3})
the inequality of the chemical potential of type 1 atoms in two phases 
($\mu_{1,1}< \mu_{1,2}$) leads to 
\beqa\label{q7}
A_\mathrm{2}n_{1,1}<g_{12}(n_{2,2}-n_{2,1}),
\eeqa
which using Eq. (\ref{q6}) yields
\beqa\label{q8}
A_\mathrm{2}^2\lambda <g_{12}^2.
\eeqa

Substituting Eq. (\ref{q6}) into Eq. (\ref{q5}) and after some 
straightforward algebra we obtain
\beqa\label{q9}
(A_\mathrm{2}^2 \lambda -g_{12}^2)(n_{2,1}-n_{2,2})^2=0,
\eeqa
which allows two possibilities. For $A_\mathrm{2}^2 \lambda  \ne 
g_{12}^2$, the 
only  solution is the trivial, nevertheless unacceptable,  one 
$n_{2,1}=n_{2,2}$, which means that 
the type 2  atoms  form a uniform configuration and not a mixed phase.  
However, if $A_\mathrm{2}^2 \lambda  = g_{12}^2$, one can have a mixed 
phase with 
$n_{2,1}\ne n_{2,2}$. Nevertheless, this condition enters in 
contradiction 
with inequality (\ref{q8}), showing that one cannot have one mixed and 
one pure phase in this case.

Next we  consider the possibility of two mixed phases. The 
equality of pressure 
and chemical potential of each 
species in two phases leads to the following conditions
\beqa
\frac{1}{2}A_\mathrm{2}(n_{1,1}^2-n_{1,2}^2)&+&\frac{1}{2}
A_\mathrm{2}\lambda 
(n_{2,1}^2-
n_{2,2}^2 ) \nonumber \\
&=&
g_{12}(n_{1,2}n_{2,2}-n_{1,1}n_{2,1}),\\
A_\mathrm{2}n_{1,1}+g_{12}n_{2,1}&=& 
A_\mathrm{2}n_{1,2}+g_{12}n_{2,2},\\
A_\mathrm{2}\lambda n_{2,1}+g_{12}n_{1,1}&= &A_\mathrm{2}\lambda 
n_{2,2}+g_{12}n_{1,2}.
\eeqa
This set of equations have only the trivial solutions 
$n_{1,1}=n_{1,2}$
and  $n_{2,1}=n_{2,2}$ corresponding to uniform mixture.
Hence two mixed phases cannot be in equilibrium.

\subsection{One-Dimensional (1D) Mixture}

From Eqs. (\ref{en-den1}), we find that 
the expression for total energy and pressure  in this case are 
\begin{eqnarray}\label{r1}
 { E}_j&=&S_j {\cal E}_j \equiv 
S_j\left[  \frac{1}{3}A_\mathrm{1}(n_{1,j}^3+\lambda 
n_{2,j}^3)+g_{12}n_{1,j}n_{2,j}\right],\nonumber \\ \\
\label{r2}
P_j&\equiv & -\frac{\partial { E}_j}{\partial S_j}=
 \frac{2}{3}A_\mathrm{1}n_{1,j}^3+g_{12}n_{1,j}n_{2,j}+
 \frac{2}{3}A_\mathrm{1}\lambda n_{2,j}^3.\nonumber \\
\end{eqnarray} 
 From Eq. 
(\ref{p1}) the chemical potentials are given by
\begin{eqnarray}\label{r3}
\mu _{1,j}&=& A _\mathrm{1}n_{1,j}^2 +g_{12}n_{2,j} \\
\label{r4x}
\mu _{2,j}&=& A_\mathrm{1}\lambda  n_{2,j}^2 +g_{12}n_{1,j}. 
\end{eqnarray}

\subsubsection{Two pure phases}
In case of two pure and separated phases one should have, 
for example,
$n_{2,1}=0$ for phase 1 and $n_{1,2}=0$ for phase 2. 
The condition of 
equal pressure then yields
\begin{eqnarray}\label{x1}
 n_{1,1}^3  = \lambda n_{2,2}^3 . 
\end{eqnarray}
For equal-mass fermions $\lambda =1$, 
and one obviously 
have have the trivial solution
$n_{1,1}=n_{2,2}$ or the  densities of the two species are equal. 
Of 
course, for $\lambda \ne 1$ the densities of the two species could 
be different. Chemical potential condition $\mu_{2,2}\le \mu_{2,1}$ 
yields
\begin{eqnarray}\label{x2}
A_\mathrm{1}\lambda n_{2,2}^2  \le g_{1,2}n_{1,1}.
\end{eqnarray}
Chemical potential condition $\mu_{1,1}\le \mu_{1,2}$ yields
\begin{eqnarray}\label{x3}
A_\mathrm{1} n_{1,1}^2  \le g_{1,2}n_{2,2}.
\end{eqnarray}
Eliminating $n_{1,1}$ between Eqs. (\ref{x1}) and (\ref{x2})
or between Eqs. (\ref{x1}) and (\ref{x3}) we get 
\begin{eqnarray}\label{x4}
n_{2,2} \le B_\mathrm{1},  \quad B_\mathrm{1}=g_{12}/(A_1\lambda^{2/3}).
\end{eqnarray}
From Eqs. (\ref{x1}) and (\ref{x4}) 
we obtain the following restriction 
on $n_{1,1}$
\begin{eqnarray}\label{x5}
n_{1,1}  \le C_\mathrm{1},  \quad 
C_\mathrm{1}=g_{12}/(A_1\lambda^{1/3}).
\end{eqnarray}

\begin{figure}[tbp]
\begin{center}
{\includegraphics[width=\linewidth,clip]{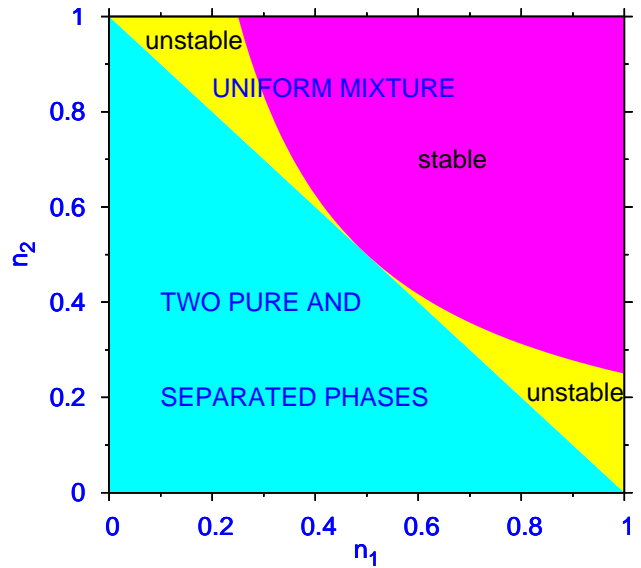}}
\end{center}
\caption{(Color online) Phase diagram for Fermi-Fermi mixture in  
one dimension (1D).  The plotted density $n_1$ is in units of 
$C_\mathrm{1}\equiv g_{12}/(A_\mathrm{1}\lambda^{1/3})$ 
and $n_2$  in 
units of
$B_\mathrm{1}\equiv g_{12}/(A_\mathrm{1}\lambda^{1/3}).$
The light gray area represents two pure and separated phases while the 
dark gray
area represents the domain of stable uniform mixture. Below  the  curved 
line in the clear area 
the uniform mixture is unstable against  small fluctuations.
}
\label{fig1}
\end{figure}

In this case  a phase diagram  showing the total 
densities of type 1 and 2 
fermions for which the system can completely separate, can be obtained 
from Eq. (\ref{q1})  
if we allow $f$ to vary from 0 to 1 and use 
conditions (\ref{x4}) and (\ref{x5}). This is illustrated in Fig. 
\ref{fig1}. 
The light gray area represents pure phases and the dark gray 
area represents the stable uniform mixture. The uniform mixture is 
unstable in the clear area
below  
the curve given by inequality  (\ref{st1}).  For attractive interaction, 
one has the formation of bright solitons by modulational instability, 
(discussed in Sec. IV). For repulsive interaction one can have a 
partially demixed 
configuration in the clear region in Fig. \ref{fig1}.

Now let us see if the system spontaneously move into the phase-separated 
configuration from an energetic consideration. The energy of the mixed 
system is 
\begin{eqnarray}\label{x7}
{ E}_{\mathrm{mix}}&=&S\left[ \frac{1}{3} A_\mathrm{1} n_1^3 
+n_1n_2g_{12}+\frac{1}{3}\lambda A_\mathrm{1} n_2^3 \right] \\
&=&S\biggr[\frac{1}{3}A_\mathrm{1}
n_{1,1}^3f^3
+ n_{1,1}n_{2,2}f(1-f)g_{1,2}\nonumber \\
&+&\frac{1}{3}
A_\mathrm{1}n_{2,2}^3 
(1-f)^3\lambda \biggr]. \label{x8}
\end{eqnarray}
Equation (\ref{x8}) is obtained with the use of Eq. (\ref{q1}).
The energy of the separated phase system with the same number of atoms 
is
\begin{eqnarray}\label{x9x}
{
E}_{\mathrm{sep}}=\frac{1}{3}A_\mathrm{1}S[
n_{1,1}^3f+n_{2,2}^3\lambda 
(1-f)
].
\end{eqnarray}
Using Eq. (\ref{x1}), 
and after some straightforward algebra, the 
difference $\Delta \equiv ({
E}_{\mathrm{mix}}-{ 
E}_{\mathrm{sep}})$ is given by 
\begin{eqnarray}\label{x9}
\Delta = n_{2,2}^2S\lambda^{1/3}
[g_{12}-A_\mathrm{1} n_{2,2}\lambda^{2/3} 
]f(1-f) 
\end{eqnarray}
Considering the restriction (\ref{x4}) 
in the separated phase, Eq. 
(\ref{x9}) yields the 
following inequality
\begin{eqnarray}\label{x10}
\Delta= n_{2,2}^2f(1-f)S\lambda^{1/3}
[1- n_{2,2}/B_\mathrm{1}
 ]g_{12}\ge 0.
\end{eqnarray}
For density ranges where equilibrium is possible $f\ne 0$ 
and $f\ne 1$, 
$ E_{\mathrm{sep}}$ is always less than $ E_{\mathrm{mix}}$. 
Hence, 
energetically the two species of fermions can separate.

\subsubsection{A mixed and a pure phase}

Now let us consider a mixed phase (phase 1) and a pure phase (phase 2) 
and consider 
the case $n_{1,2}=0$. The equality of pressure now leads to 
\begin{eqnarray}\label{ab1}
\frac{2}{3}A_\mathrm{1}n_{1,1}^3 + \frac{2}{3}A_\mathrm{1} 
\lambda n_{2,1}^3+g_{12}n_{1,1}n_{2,1}
= \frac{2}{3}A_\mathrm{1} \lambda n_{2,2}^3.  
  \end{eqnarray}
The equality of chemical potential of species 2  in 
two phases ($\mu_{2,1}=\mu_{2,2}$) yields 
\begin{eqnarray}\label{ab2}
n_{1,1}=A_\mathrm{1}\lambda 
(n_{2,2}^2-n_{2,1}^2)/ g_{12} . 
  \end{eqnarray}
Eliminating $n_{1,1}$ between equations (\ref{ab1}) and (\ref{ab2}) 
(after some straightforward algebra)
we 
get
\begin{eqnarray}\label{ab3}
2\Lambda (1-x^2)^3=x^3-3x+2
  \end{eqnarray}
where $x=n_{2,1}/n_{2,2}$, $\Lambda=  (n_{2,2}/B_\mathrm{1})^3$. After 
cancelling the 
trivial factor $(1-x)^2$ from both sides of Eq. (\ref{ab3}), we get
\begin{eqnarray}\label{ab4}
2\Lambda (1+x)^3(1-x)=x+2
  \end{eqnarray}
From Eq. (\ref{ab4}) we find that the solution $x=0$ is obtained for 
$\Lambda =1$  corresponding to $n_{2,1}= n_{1,2}=0$, 
$n_{2,2}=B_\mathrm{1}$ and $n_{1,1}=C_\mathrm{1}$. 
The densities of the first component are $n_{1,2}=0$ and
$n_{1,1}= C_\mathrm{1}.$  This is the special case considered in Sec. IIIB1
[see, Eqs. (\ref{x4}) and (\ref{x5})]. 
The solution $n_{2,2}=B_\mathrm{1}$ $ 
(\Lambda =1)$ is a solution of two pure phases corresponding to 
$x=0$.
 The domain of solution of mixed phase corresponds to 
$n_{2,2}>B_\mathrm{1} $ $
(\Lambda > 1)$ corresponding to $x>0$ (recall that the fraction $x$ 
cannot be negative.) 
Hence for the present mixed phase to exist Eq. 
(\ref{ab4}) should have the solution $x\to +0$ for $\Lambda \to +1$. 
However, we find from Eq. (\ref{ab4}) as $\Lambda$ is made slightly 
greater than 1, the solution $x=0$ turns negative (unphysical). [Please 
note that for $\Lambda=1$ Eq. (\ref{ab4}) has two real roots: 
$x=0,$ $0.7399...$; the latter (spurious)
root is not of present physical interest.] 
Hence we conclude that a mixed and a pure phase cannot be realized in 
the present mixture.    

Finally, one can consider the possibility of two mixed phases. The 
equality of pressure and chemical potential
  of each 
species in two phases  leads to
\beqa
\frac{2}{3}A_\mathrm{1}(n_{1,1}^3-n_{1,2}^3)&+&\frac{2}{3}A_\mathrm{1}\lambda 
(n_{2,1}^3-
n_{2,2}^3 ) \nonumber \\
&=&
g_{12}(n_{1,2}n_{2,2}-n_{1,1}n_{2,1}),\\
A_\mathrm{1}n_{1,1}^2+g_{12}n_{2,1}&=& 
A_\mathrm{1}n_{1,2}^2+g_{12}n_{2,2},\\
A\lambda n_{2,1}^2+g_{12}n_{1,1}&= &A\lambda n_{2,2}^2+g_{12}n_{1,2}.
\eeqa
This set of equations have only the trivial solutions 
$n_{1,1}=n_{1,2}$
and  $n_{2,1}=n_{2,2}$ corresponding to uniform mixture
and that is also possible when the condition of uniform mixture is 
satisfied. Hence two mixed phases cannot be in equilibrium.

\subsection{Three-dimensional (3D) Mixture}

From Eqs. (\ref{en-den3}), we find that 
the expression for total energy and pressure  in this case are 
\begin{eqnarray}\label{r13}
 { E}_j&=&S_j {\cal E}_j \equiv 
S_j\left[  \frac{3}{5}A_\mathrm{3}n_{1,j}^{5/3}+g_{12}n_{1,j}n_{2,j}+
 \frac{3}{5}A_\mathrm{3}\lambda n_{2,j}^{5/3} \right],\nonumber \\ \\
\label{r23}
P_j&\equiv & -\frac{\partial { E}_j}{\partial S_j}=
 \frac{2}{5}A_\mathrm{3}n_{1,j}^{5/3}+g_{12}n_{1,j}n_{2,j}+
 \frac{2}{5}A_\mathrm{3}\lambda n_{2,j}^{5/3}.\nonumber \\
\end{eqnarray} 
 From Eq. 
(\ref{p1}) the chemical potentials are given by
\begin{eqnarray}\label{r3x}
\mu _{1,j}&=& A_\mathrm{3} n_{1,j}^{2/3} +g_{12}n_{2,j} \\
\label{r4}
\mu _{2,j}&=& A_\mathrm{3}\lambda  n_{2,j}^{2/3} +g_{12}n_{1,j}. 
\end{eqnarray}

\subsubsection{Two pure phases}

Again for two pure and separated phases we take 
 $n_{1,2}=n_{2,1}=0$. The condition of equal pressure in two phases 
then leads to 
\beq  \label{k1}
n_{1,1}^{5/3}=\lambda n_{2,2}^{5/3}.
\eeq
The chemical potential condition $\mu_{2,2} \le \mu_{2,1}$ yields
\beq  \label{k2}
n_{1,1}\ge \lambda A_\mathrm{3}n_{2,2}^{2/3}/g_{12}.
\eeq
 The chemical potential condition $\mu_{1,1} \le \mu_{1,2}$ yields
\beq  \label{k3}
n_{2,2}\ge A_\mathrm{3}n_{1,1}^{2/3}/g_{12}.
\eeq
Eliminating $n_{1,1}$ between Eqs. (\ref{k1}) and (\ref{k2})
or between Eqs. (\ref{k1}) and (\ref{k3}) we obtain 
\beq  \label{k4x}
n_{2,2}\ge B_\mathrm{3}, \quad B_\mathrm{3}=(\lambda 
^{2/5}A_3/g_{12})^3.
\eeq
Similarly, eliminating $n_{2,2}$ 
between Eqs. (\ref{k1}) and (\ref{k2}) we get 
\beq  \label{k4}
n_{1,1}\ge C_\mathrm{3}, \quad C_\mathrm{3}=(\lambda 
^{3/5}A_3/g_{12})^3.
\eeq

\begin{figure}[tbp]
\begin{center}
{\includegraphics[width=\linewidth,clip]{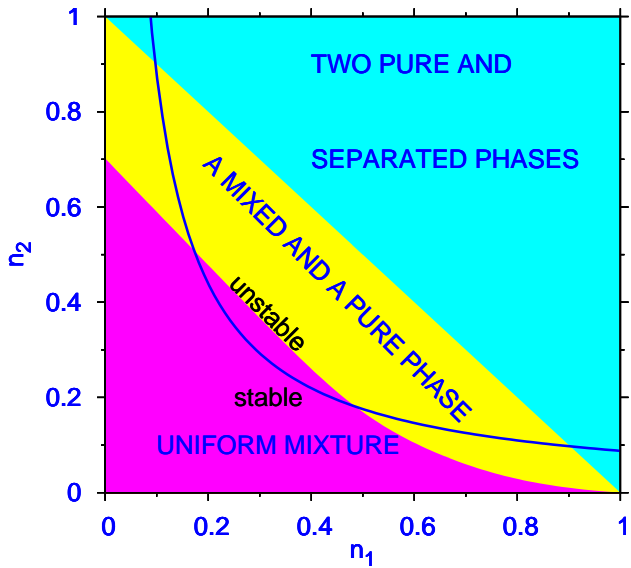}}
\end{center}
\caption{(Color online) Phase diagram for Fermi-Fermi mixture in  
three dimensions (3D).  The plotted density $n_1$ is in units of 
$C_\mathrm{3}\equiv  (\lambda^{2/5}A_\mathrm{3}/g_{12})^3$ 
and $n_2$  in units of
$B_\mathrm{3} \equiv (\lambda^{3/5}A_\mathrm{3}/g_{12})^3.$
The light gray area represents two pure and separated phases while the 
dark gray
area represents the domain of stable uniform mixture. The clear area 
represents the region where a mixed and a pure phase can exist. 
Above  the  curved 
line 
the uniform mixture is unstable against small fluctuations.
}
\label{fig2}
\end{figure}

In this case  a phase diagram  showing the total
densities of type 1 and 2
fermions for which the system can completely separate, 
can be obtained
from Eq. (\ref{q1})
if we allow $f$ to vary from 0 to 1 and use
conditions (\ref{k4x}) and (\ref{k4}). This is illustrated in Fig.
\ref{fig2}.

To see the separation of the two types of fermions from an energetic 
consideration, we calculate the energies of the mixed and separated 
configurations. The energy of the mixed phase is \cite{viverit}
\beqa
{ E}_{\mathrm{mix}}&=& S\left[ 
\frac{3}{5}A_\mathrm{3}n_1^{5/3}+g_{12}n_1n_2
+\frac{3}{5}A_\mathrm{3}\lambda n_2^{5/3}\right] \\
&=& S\biggr[ 
\frac{3}{5}A_\mathrm{3}f^{5/3}n_{1,1}^{5/3}+g_{12}n_{1,1}n_{2,2}f(1-f)\nonumber 
\\
&+& \frac{3}{5}A_\mathrm{3}\lambda n_{2,2}^{5/3}(1-f)^{5/3}\biggr]. 
\eeqa
The energy of the separated phase is 
\beqa
{ E}_{\mathrm{sep}}&=& S\left[\frac{3}{5}A_\mathrm{3}n_{11}^{5/3}f
+\frac{3}{5}A_\mathrm{3}\lambda n_{2,2}^{5/3}(1-f) \right]. 
\eeqa
Using Eq. (\ref{k1}) the difference $\Delta =( {\cal 
E}_{\mathrm{mix}}-{\cal E}_{\mathrm{sep}})$ can be written as
\beqa \label{k6}
\Delta &=& S[3A_3\lambda 
n_{2,2}^{5/3}f(f^{2/3}-1)/5+g_{12}n_{22}^2\lambda^{3/5}f(1-f)\nonumber 
\\
&+& 3A_\mathrm{3}\lambda n_{2,2}^{5/3}(1-f)\{(1-f)^{2/3}-1\}/5].
\eeqa
Using inequality (\ref{k3}), Eq. (\ref{k6}) yields
\beqa \label{k7}
\Delta &\ge & 
\frac{A_\mathrm{3}^6\lambda^3}{g_{12}^5}\biggr[-\frac{3}{5}f(1-f^{2/3}) 
\nonumber 
\\
&+& f(1-f)-\frac{3}{5}(1-f)\{1-(1-f)^{2/3}\}  \biggr]. 
\eeqa
For $1>f>0$, the quantity given by (\ref{k7}) is always positive. Hence 
the separated phase has less energy than the mixed phase and the system 
will spontaneously move into the phase separated configuration. 

In this case also two mixed phases cannot be in equilibrium as in 1D.

\subsubsection{A mixed and a pure phase}

Again we consider a mixed (species 2) and a pure  (species 1) phase 
and consider the case  $n_{1,2}=0$. The equality of pressure now leads 
to 
\beqa  \label{m1}
\frac{2}{5}A_\mathrm{3}n_{1,1}^{5/3}+g_{12}n_{1,1}n_{2,1}+ 
\frac{2}{5}A_\mathrm{3}\lambda n_{2,1}^{5/3}=
\frac{2}{5}A_\mathrm{3}\lambda n_{2,2}^{5/3}.
\eeqa
The equality of chemical potential of species 2 in two phases ($\mu 
_{2,1}= \mu_{2,2}$) yields
\beqa \label{m2}
n_{1,1}= A_\mathrm{3}\lambda (n_{2,2}^{2/3}-n_{2,1}^{2/3})/g_{12}.
\eeqa
Eliminating $n_{1,1}$ between Eqs. (\ref{m1}) and (\ref{m2}) and after 
some straightforward algebra we get 
\beqa \label{m3}
2\Lambda (1+x)^{5/3}-(1-x)^{1/3}(3x^3+6x^2+4x+2)=0, 
\eeqa
where $x=(n_{2,1}/n_{2,2})^{1/3}$, and $\Lambda = 
(n_{2,2}/B_\mathrm{3})^{-5/9}$. 
From Eq. (\ref{m3}) we find that the solution $x=0$ is obtained for 
$\Lambda =1$ corresponding to $n_{2,1}=0$,  $n_{2,2}=B_\mathrm{3}$,  
$n_{1,1}=C_\mathrm{3}$, $n_{1,2}=0$. This is the limiting case of two
pure and separated phase studied in Sec. IIIC1.  
(In addition for $\Lambda =1$, Eq. (\ref{m3}) has the spurious 
or unphysical root $x=0.90278...$, which we do not consider here.) 
For two purely separated phases we have seen that $n_{2,2}\ge 
B_\mathrm{3}$ whence 
$\Lambda \equiv (n_{2,2}/ B_\mathrm{3})^{-5/9} \le 1.$ The domain for a 
mixed and a 
separated phase then should have $\Lambda > 1$. 
To find this domain we solve Eq. (\ref{m3}) for $x>0$ using different 
$\Lambda$. Such solutions appear  in the range $1.217... \ge \Lambda \ge 
1$. 
Using this solution for $x$
we obtain $n_{2,2}$ and  $n_{2,1}$
from the definitions of  $\Lambda$ and $x$, respectively. Finally, 
$n_{1,1}$ is obtained from Eq. (\ref{m2}).
The results so-obtained for $n_{2,2}$, $n_{2,1}$, and $n_{1,1}$ 
for different $\Lambda$
are used in 
\begin{equation}
n_1=fn_{1,1}, \quad \mbox{and}  \quad n_2=fn_{2,1}+(1-f)n_{2,2},
\end{equation}   
to calculate the domain of $n_1$ and $n_2$, by varying $f$ in the range 
$1>f>0$, 
which allows a pure and a 
mixed phase.

We show the 3D phase diagram 
for 
total densities 
of type 1 and 2 fermions 
in Fig. \ref{fig2}. 
In this figure the light gray  area represents the domain of 
two separated 
phases 
and  the clear area  that of  a 
mixed and a separated phase as calculated above.
The remaining dark
gray area represents the domain of  stable uniform mixture. 
The uniform mixture is unstable  above
the curve given by Eq. (\ref{st3}).  Qualitatively, Fig. \ref{fig2} is 
quite similar to Fig. 3  of Viverit {\it et al.} \cite{viverit} for a 
Bose-Fermi mixture.

If we compare Figs.  \ref{fig1} and  \ref{fig2}
 we find that in 1D the pure phases appear at small densities, and 
uniform mixture at large densities. The uniform mixture is stable at 
larger densities. The opposite happens in 3D. If we compare the findings 
of Viverit {\it et al.} \cite{viverit} for a study of the phase diagram 
of a 
Bose-Fermi mixture in 3D and compare 
with the study of Das \cite{das} in 1D we find that such an inversion 
also takes 
place there.   Moreover in 1D there cannot be a mixed and a pure phase 
for a Fermi-Fermi mixture, 
which is possible in 3D.


\section{Dynamical Equations in Quasi-1D Superfluid Fermi-Fermi Mixture}

\subsection{The Model}

Of the three dimensional possibilities $-$ 1D, 2D and 3D $-$ the 1D case 
deserves special attention. In 1D, if the 
interspecies Fermi-Fermi 
interaction is attractive, in the domain of instability of the uniform 
mixture one can have the formation of bright soliton by modulational 
instability. To perform a careful study of  the nature of these bright 
solitons (and their 
dynamical stability)  we derive the Euler-Lagrange equations 
in 1D from its Lagrangian density.

We consider a mixture of $N_1$ superfluid atomic fermions   
of mass $m_1(=m)$ and $N_2$ superfluid  
atomic fermions of mass $m_2 (=m/\lambda)$ at zero temperature 
trapped by a tight cylindrically symmetric harmonic 
potential of frequency $\omega_{\bot}$ in the 
transverse (radial cylindric) direction. We assume factorization 
of the transverse degrees of freedom. 
This is justified in 1D confinement where, regardless
of the longitudinal behavior
or statistics, the transverse spatial profile is that
of the single-particle ground-state \cite{das,sala-st,sala-npse}. 
The transverse width of the atom distribution  
is given by the characteristic harmonic length of the 
single-particle ground-state: 
$a_{\bot j}=\sqrt{\hbar/(m_j\omega_{\bot})}$, with $j=1,2$. 
The atoms have an effective 1D 
behavior at zero temperature if their 
chemical potentials are much smaller than 
the transverse energy $\hbar\omega_{\bot}$ \cite{das,sala-st,sala-npse}. 
The interspecies Fermi-Fermi interaction is characterized 
by a contact potential with 
scattering length $a_{12}$, which can be repulsive 
or attractive.

We use a mean-field Lagrangian 
to study the static and collective properties 
of the 1D superfluid Fermi-Fermi mixture as in the Ginzburg-Landau 
theory \cite{fw}. 
The Lagrangian density ${\cal L}$ of the mixture reads 
\beq 
{\cal L} = {\cal L}_1 + {\cal L}_2 + {\cal L}_{12} \; . 
\label{l-tot}
\eeq 
The term ${\cal L}_i$ is the fermionic  Lagrangian for component $i$, 
defined as 
\beqa
{\cal L}_i  = {i\hbar\over 2}\, \left(\psi_i^* 
{\partial \psi_i\over 
\partial 
t}  - \psi_i {\partial \psi_i^*\over \partial
t}
\right)
- {\hbar^2   \over 2{m_{\mathrm{eff}}^{(i)}}       }
 \left| {\partial \psi_i 
\over \partial z} 
\right|^2 
-\frac{A_{\mathrm{1}}^{(i)}}{3} |\psi_i|^6, \nonumber \\
\label{l-b}
\eeqa
where $A_{\mathrm{1}}^{(i)}=\hbar^2\pi^2/(8m_i)$,
$\psi_i(z,t)$ is the field of the $i$th component
of the BCS Fermi superfluid
along the longitudinal axis, 
such that $n_i(z,t)=|\psi_i(z,t)|^2$ is the 1D local probability 
density of  the $i$th component.
Here $m_{\mathrm{eff}}^{(i)}$ is the effective mass of 
superfluid flow 
in the Ginzburg-Landau theory. There is 
experimental evidence \cite{fw} that this effective mass is twice the 
fermion mass
($m_{\mathrm{eff}}^{(i)}=2 m_i$) and we shall use this effective mass in 
the following study.

Finally, the Lagrangian density ${\cal L}_{12}$ 
of the interaction between the two Fermi components is 
taken to be of the following standard 
zero-range form \cite{sala-st,sala-sadhan2}
\beq 
{\cal L}_{12} = - g_{12} \, 
|\psi_1|^2 |\psi_2|^2 \; ,  
\label{l-bf} 
\eeq 
where $g_{12}=2 \hbar\omega_{\bot} a_{12}$ 
is the 1D Fermi-Fermi interaction strength.

The Euler-Lagrange equations of the Lagrangian
${\cal L}$ are the two following
coupled partial differential equations:
\beqa
i\hbar \partial_t \psi_1 &=& \left[ -\frac{\hbar^2}{4m_1} \partial_z^2
+ A_{\mathrm{1}}^{(1)}n_1^2
 + g_{12} n_2 \right] \psi_1  \; ,
\label{str1a}\\
i \hbar \partial_t \psi_2 &=& \left[ - \frac{\hbar^2}{4m_2} 
\partial_z^2
+ A_{\mathrm{1}}^{(2)}   n_2^2
 + g_{12} n_1 \right] \psi_2  \; ,
\label{str2a}
\eeqa
with the normalization $\int_{-\infty}^\infty |\psi_i|^2 dz = N_i$.

{It is convenient to work in terms of dimensionless 
variables defined in 
terms of a frequency $\omega$ and length  
$l\equiv \sqrt{\hbar / (2m_1 \omega)}$ by 
$\psi_j=\hat \psi_j/\sqrt l$, 
$t= 2 \hat t/\omega$,  $z=\hat z l$, 
and $g_{12}={\hat g_{12} \hbar^2/(4 m_1l)}$. With  these new 
variables Eqs. (\ref{str1a}) and (\ref{str2a})
can be written as
\beqa
i \partial_t \psi_1 &=& \biggr[ - \partial_z^2
+ An_1^2
 + g_{12} n_2 \biggr] \psi_1  \; ,
\label{str1}\\
i \partial_t \psi_2 &=& \biggr[ - \lambda  \partial_z^2
+\lambda A n_2^2
 + g_{12} n_1 \biggr] \psi_2  \; ,
\label{str2}
\eeqa
where $A \equiv \pi^2/2$ and where 
we have dropped the hats over the variables, and where  
$\lambda= m_1/m_2,$  
$n_i=|\psi_i|^2, i=1,2$ 
with the normalization $\int_{-\infty}^\infty n_i dz = N_i$. 
Equations (\ref{str1}) and (\ref{str2}) with diagonal quintic 
nonlinearity are the equations satisfied by two coupled TG 
Bose gas \cite{Tonks} and hence the analysis of Sec. IV also 
applies to a TG
gas. 
}

For stationary states the solution of Eqs. (\ref{str1}) and (\ref{str2})
have the form $\psi_i=\phi_i \exp(-i\mu_it)$ where $\mu_i$ are the 
respective chemical potentials. Consequently, these equations reduce to  
\beqa
\mu_1 \phi_1 &=& \biggr[ - \partial_z^2
+ An_1^2
 + g_{12} n_2 \biggr] \phi_1  \; ,
\label{mu1}\\
\mu_2 \phi_2 &=& \biggr[ - \lambda  \partial_z^2
+\lambda A n_2^2
 + g_{12} n_1 \biggr] \phi_2  \; ,
\label{mu2}
\eeqa
A repulsive interspecies Fermi-Fermi 
interaction is produced by a 
positive 
$g_{12}$, while an attractive Fermi-Fermi interaction 
corresponds to a negative $g_{12}$.

\subsection{Modulational Instability}

To study analytically the modulational instability 
\cite{ff,sala-prl} 
of Eqs. (\ref{str1}) and (\ref{str2}) 
we consider the special 
case of attractive Fermi-Fermi 
interaction  while these equations reduce to  
\beqa
i \partial_t \psi_1 &=& \left[ - \partial_z^2
+ A  |\psi_1|^4   - g_{12} |\psi_2|^2 \right] \psi_1  \; ,
\label{m1x}\\
i \partial_t \psi_2 &= &\left[ - \lambda \partial_z^2
+A\lambda  |\psi_2|^4
 - g_{12} |\psi_1|^2 \right] \psi_2  \; ,
\label{m2x}
\eeqa
where    
we have taken the interspecies interaction to be attractive 
by inserting an explicit negative sign in  $g_{12}$.

We analyze the modulational instability of a 
constant-amplitude solution corresponding to a uniform mixture 
in coupled Eqs.  (\ref{m1x}) and (\ref{m2x}) by considering 
the solutions
\begin{eqnarray}
\varphi_{10}=\aleph_{10}\exp(i\delta_1)\equiv \aleph_{10}e^{i t (g_{12}
\aleph_{20}^2
-A  \aleph_{10}^4 )}, \label{s1}
\end{eqnarray}
\begin{eqnarray}
\varphi_{20}=\aleph_{20}\exp(i\delta_2)\equiv \aleph_{20}e^{i t (g_{12}
\aleph_{10}^2
-\lambda A \aleph_{20}^{4} )}, \label{s2}
\end{eqnarray}
of  Eqs.  (\ref{m1x}) and (\ref{m2x}), respectively, where $\aleph_{i0}$ 
is the 
amplitude and 
$\delta_i$ a phase
for component $i$. The constant-amplitude solutions, describing an 
uniform mixture, 
develop an
amplitude-dependent phase on time evolution.
We consider a  small perturbation $\aleph_i\exp(i\delta_i)$ to
these solutions via
\begin{eqnarray}
\varphi_i=(\aleph_{i0}+\aleph_i)\exp(i\delta_i),
  \end{eqnarray}
where $\aleph_i=\aleph_i(z,t)$.
Substituting these perturbed solutions in Eqs.  (\ref{m1x}) and  
(\ref{m2x}),
and for small perturbations retaining
only the linear terms in $\aleph_i$ we get
\begin{eqnarray}\label{q1x}
&i&{\partial_t  \aleph_1}
+ {\partial_z ^2 \aleph_1} 
-    2A \aleph_{10}^{4}(\aleph_1+\aleph_1^*)\nonumber \\
&+&
g_{12}
\aleph_{10}\aleph_{20} (\aleph_2+\aleph_2^*)=0, \\
\label{q2}
&i&{\partial_t \aleph_2} + \lambda {\partial_z ^2 \aleph_2} 
-   2\lambda A \aleph_{20}^{4}(\aleph_2+\aleph_2^*)\nonumber \\
&+&g_{12}
\aleph_{10}\aleph_{20} (\aleph_2+\aleph_2^*)=0. 
\end{eqnarray}
We consider the complex plane-wave
perturbation
\begin{equation}
\aleph_i(z,t)=
{\cal A}_{i1} \cos (Kt -\Omega z)+i {\cal A}_{i2} \sin  (Kt 
-\Omega
z) \label{comp}
\end{equation}
with $i=1,2$, where ${\cal A}_{i1}$ and  
${\cal A}_{i2}$
are the amplitudes for the real and imaginary parts, respectively, and
$K$ and $\Omega$ are frequency and wave numbers.

Substituting  Eq. (\ref{comp}) in Eqs. (\ref{q1x}) and (\ref{q2}) and  
separating the real and imaginary parts we get
\begin{eqnarray}\label{p1x}
-{\cal A}_{11}K&=&{\cal A}_{12}\Omega^2, \\
-{\cal
A}_{12}K&=&{\cal
A}_{11}\Omega^2-2g_{12}\aleph_{10}\aleph_{20}{\cal A}_{21}+
4A
\aleph_{10}^{4}{\cal   A}_{11}, \nonumber \\ \label{p2x}
  \end{eqnarray}
for  $i=1$, and
\begin{eqnarray}\label{p3x}
-{\cal   A}_{21}K&=&{\cal   A}_{22}\Omega^2\lambda, \\
-{\cal
A}_{22}K&=&{\cal
  A}_{21}\Omega^2\lambda-2g_{12}\aleph_{10}\aleph_{20}{\cal 
A}_{11}+4A\lambda 
\aleph_{20}^{4}{\cal   A}_{21},   \nonumber \\  \label{p4x}
 \end{eqnarray}
for $i=2$.
Eliminating ${\cal A}_{12}$ between Eqs.  (\ref{p1x})  and
(\ref{p2x}) we get 
\begin{eqnarray}\label{p5x}
{\cal   A}_{11}[K^2-\Omega^2( \Omega^2+
4A  \aleph_{10}^4
)]=-
2{\cal A}_{21}g_{12}\aleph_{10}\aleph_{20} \Omega^2, \nonumber \\ 
\end{eqnarray}
and eliminating ${\cal A}_{22}$ between  (\ref{p3x})  and
(\ref{p4x}) we have 
\begin{eqnarray}\label{p6x}
 {\cal A}_{21}[K^2-\Omega^2\lambda( \Omega^2\lambda+4A\lambda 
\aleph_{20}^{4})] =-
2{\cal A}_{11}g_{12}\aleph_{10}\aleph_{20} \Omega^2\lambda.  \nonumber 
\\
\end{eqnarray}
Finally, eliminating ${\cal A}_{11}$ and ${\cal A}_{21}$
from   (\ref{p5x}) and (\ref{p6x}) and recalling that the density of the 
uniform mixture $n_1$ and $n_2$ of the two species are given by $n_i=
\aleph_{i0}^2$,
we obtain the following
dispersion relation
\begin{eqnarray}\label{p7x}
&2K&= \pm \Omega[ \left(\Omega^2+\Omega^2\lambda+
4A n_1^2
+
4A \lambda^2 n_2^2\right) 
\pm \{ ( \Omega^2
\nonumber \\
&-&\Omega^2\lambda^2
+ 4A n_1^2
-4A \lambda^2 n_2^2 )^2+
16g_{12}^2\lambda n_1 n_2
 \}^{1/2} ]^{1/2}.\nonumber \\
\end{eqnarray}

For stability of the plane-wave perturbation,  $K$ has to be
real. For any $\Omega$ this happens for
\begin{eqnarray}
(4A n_1^2
+4A\lambda^2  n_2^2)^2 
> \left(
4A n_1^2 -
4A \lambda^2 n_2^2 \right)^2 +
16g_{12}^2
n_1n_2\lambda, \nonumber \\
\end{eqnarray}
or for 
\begin{equation}\label{con3}
4A^2
 \lambda  n_1n_2 \ge g_{12}^2.
\end{equation}
However, for
$4A^2
 \lambda  n_1n_2 < g_{12}^2$,
$K$ can become imaginary and the plane-wave perturbation 
can grow exponentially with time. This is the domain of modulational
instability of a constant-amplitude solution (uniform mixture) 
signalling the possibility
of coupled Fermi-Fermi bright soliton to appear
[compare with inequality  (\ref{st1}) of Sec. IIB describing stability 
of 
uniform mixture. The transformation of the quantities in inequality  
(\ref{st1}) to the dimensionless variables of inequality 
(\ref{con3}) can be performed with the definitions given after Eq. 
(\ref{str2a}).]

\subsection{Variational Results}

Here we develop a variational localized solution to Eqs. 
(\ref{mu1}) and (\ref{mu2}) 
noting that these equations can be derived from the Lagrangian 
\cite{VA} 
\beqa
\label{lag}
L&=&\int_{-\infty}^\infty\biggr[ \mu_1\phi_1^2
+\mu_2\phi_2^2-(\phi_1') ^2- (\phi_2')^2
- \phi_1^6 A/3
\nonumber \\ 
&-&\lambda \phi_2^6 A/3-g_{12}\phi_1^2\phi_2^2\biggr]
dz -\mu_1N_1 - \mu_2N_2
\eeqa
by demanding $\delta L/\delta \phi_1=\delta L/\delta 
\phi_2=\delta L/\delta \mu_1=\delta L/\delta \mu_2=0$.

To develop the variational approximation we use the following 
Gaussian ansatz \cite{sala-variational} 
\beqa
\phi_1(z)&=&\pi ^{-1/4}\sqrt{\frac{N_1 \alpha_1}{w_1}}
\exp \left(
-\frac{z^{2}}{2w_1^{2}}\right) ,\\
\phi_2(z)&=&\pi ^{-1/4}\sqrt{\frac{N_2 \alpha_2}{w_2}}
\exp \left(
-\frac{z^{2}}{2w_2^{2}}\right) ,
\label{Gauss}
\eeqa
where the variational parameters are $\alpha_j$, 
the solitons' norm, and $w_j$ 
width, in addition to $\mu_j$. 
The substitution of this variational ansatz in Lagrangian 
(\ref{lag}) yields 
\beqa
L&=&\mu_1 N_1
(\alpha_1-1)+\mu_2N_2(\alpha_2-1)-\frac{N_1\alpha_1}{2w_1^{2}}-
\frac{N_2\alpha_2}{2w_2^{2}}
\nonumber \\
&-&\frac{A\alpha _1^3N_1^3}{3\pi \sqrt 3 
w_1^2}
-\frac{A\lambda \alpha_2^3N_2^3}{3\sqrt 3 \pi w_2^2}
-\frac{g_{12}N_1N_2\alpha_1\alpha_2}{\sqrt{\pi(w_1^2+w_2^2)}}
.  \label{LG}
\eeqa

The first variational equations emerging from Eq.
(\ref{LG}) 
$\partial L/\partial \mu_1=\partial L/\partial \mu_2=0$
yield $\alpha_1=\alpha_2=1$. Therefore the conditions 
$\alpha_1=\alpha_2=1$
will be substituted in the subsequent variational 
equations. The variational equations 
$\partial L/\partial 
w_j=0$ lead to 
\beqa
&1&+ \frac{2N_1^2A}{3\pi\sqrt 3}
+\frac{g_{12}N_2w_1^4}{\sqrt\pi  
(w_1^2+w_2^2)^{3/2}}=0,  \label{WG1}\\
&1&+\frac{2N_2^2A\lambda}{3\sqrt 3 \pi}
+\frac{g_{12}N_1w_2^4}{\sqrt\pi  
(w_1^2+w_2^2)^{3/2}}
=0.  \label{WG2}
\eeqa

The remaining variational equations  are
$\partial L/\partial \alpha_j=0$, which
yield $\mu $ as a function of $w_j$'s,  and $g$'s:
\beqa
\mu_1 &=& \frac{1}{2w_1^{2}}
+
\frac{\sqrt 3N_1^2A}{3\pi w_1^2}
+\frac{g_{12}N_2}{\sqrt{\pi(w_1^2+w_2^2)}}.  \label{muG1}\\
\mu_2 
&=& \frac{1}{2w_2^{2}}+
\frac{\sqrt 3N_2^2\lambda A}{3\pi w_2^2}
+\frac{g_{12}N_1}{\sqrt{\pi(w_1^2+w_2^2)}}.  
\label{muG2}
\eeqa
Equations (\ref{WG1}) $-$ (\ref{muG2}) are the variational results which 
we shall use in our study of bright Fermi-Fermi solitons. 

\begin{figure}[tbp]
\begin{center}
{\includegraphics[width=\linewidth]{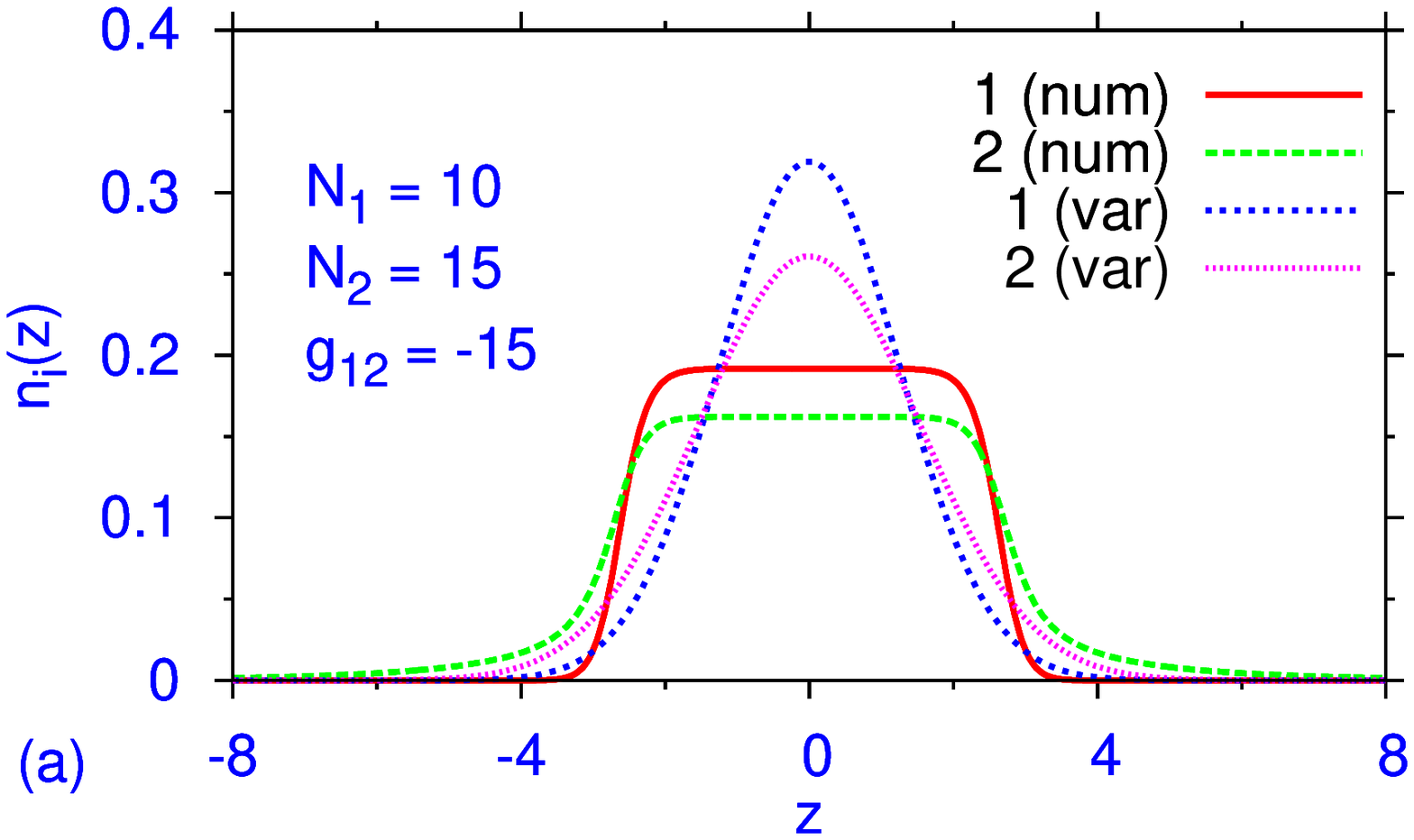}}
{\includegraphics[width=\linewidth]{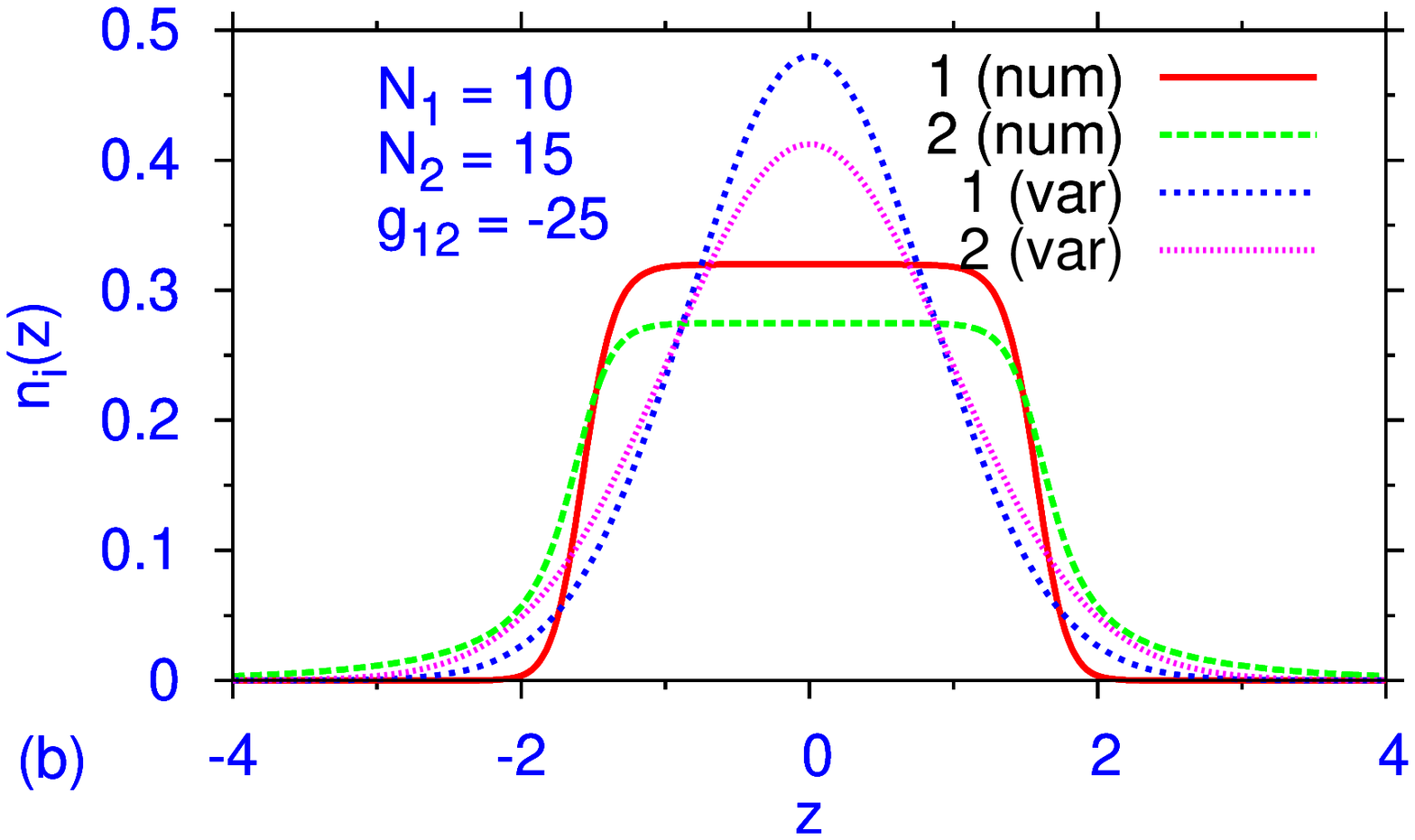}}
{\includegraphics[width=\linewidth]{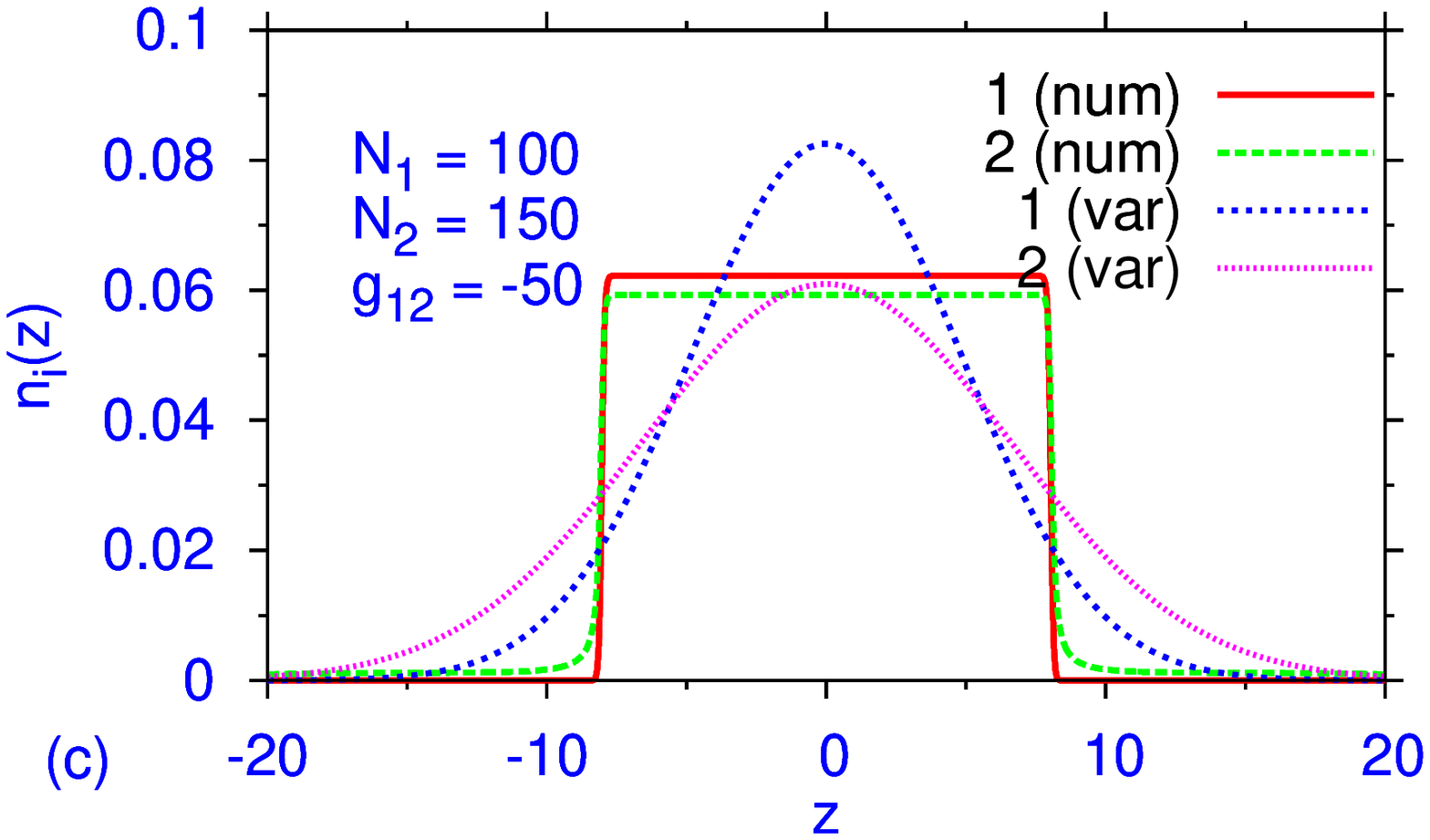}}
{\includegraphics[width=\linewidth]{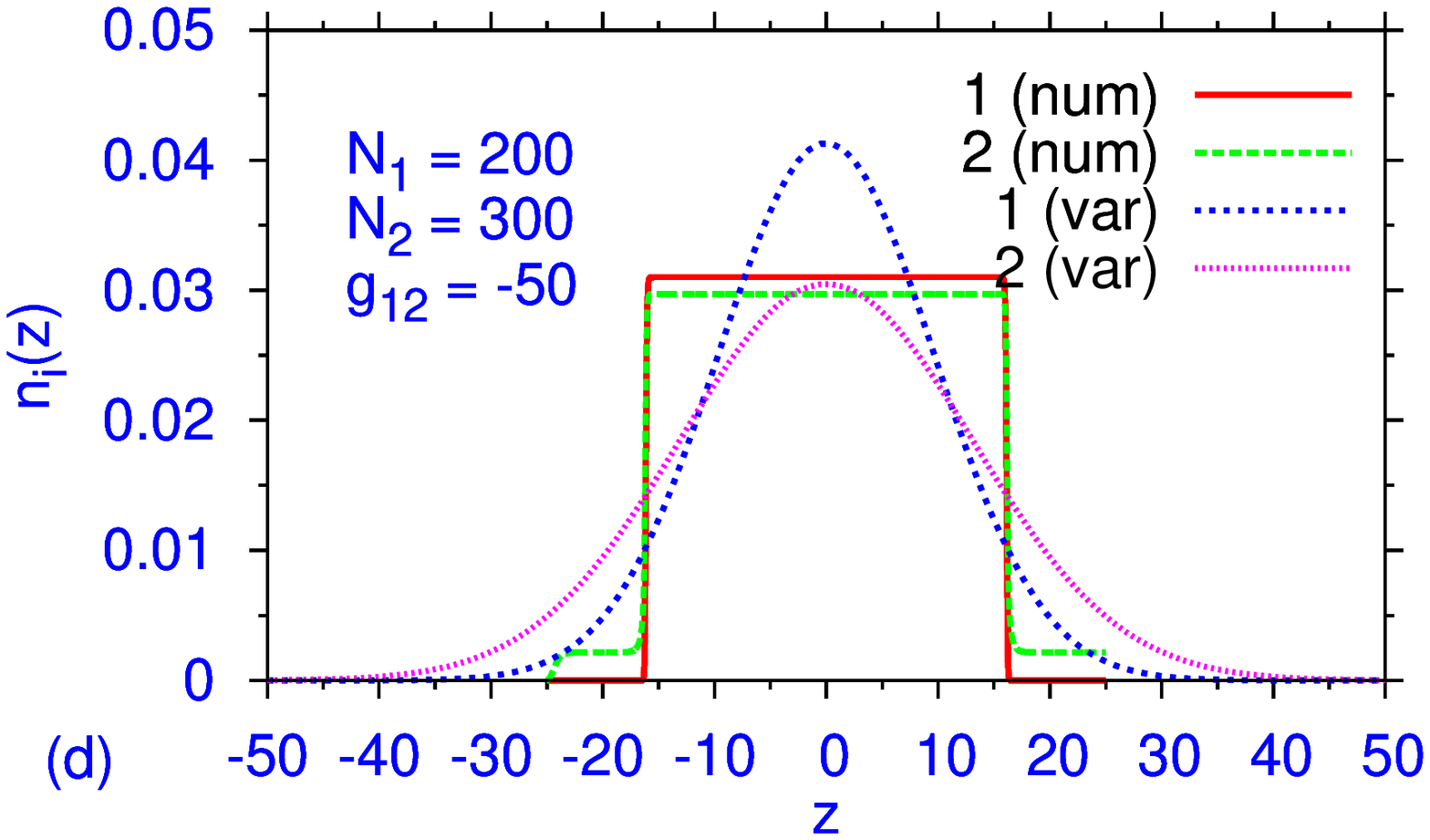}}
\end{center}
\caption{(Color online)
Probability densities of the two fermion components 
from the numerical 
solution 
(labelled ``num")
of Eqs. (\ref{mu1}) and (\ref{mu2})
  (here normalized to 
unity: $\int_{-\infty}^\infty n_i(z) dz=1 $)
compared with
variational results 
(labelled ``var")
given by Eqs.  (\ref{WG1}) and (\ref{WG2}) for $\lambda =1$ and 
 (a) $N_1=10, N_2=15$ and $g_{12}= -15$,
 (b) $N_1=10, N_2=15$ and $g_{12}= -25$,
 (c) $N_1=100, N_2=150$ and $g_{12}= -50$, and
 (d) $N_1=200, N_2=300$ and $g_{12}= -50$.
}
\label{fig3}
\end{figure}

\subsection{Numerical Results}

For stationary solutions we solve time-independent 
Eqs. (\ref{mu1}) and (\ref{mu2}) by using an 
imaginary time propagation method based on the finite-difference 
Crank-Nicholson discretization scheme of time-dependent 
Eqs. (\ref{str1}) and (\ref{str2}). The non-equilibrium 
dynamics from an initial stationary state
is studied by solving the time-dependent Eqs. (\ref{str1}) and
(\ref{str2}) with real time propagation by using as initial input the  
solution obtained by the imaginary time propagation method. The reason 
for this mixed treatment is that the imaginary time propagation method 
deals with real variables only and provides very accurate 
solution of the stationary problem at low computational 
cost \cite{muru}. In the finite-difference discretization we use space 
step of 0.025 and time step of 0.0005. 

\begin{figure}[tbp]
\begin{center}
{\includegraphics[width=\linewidth]{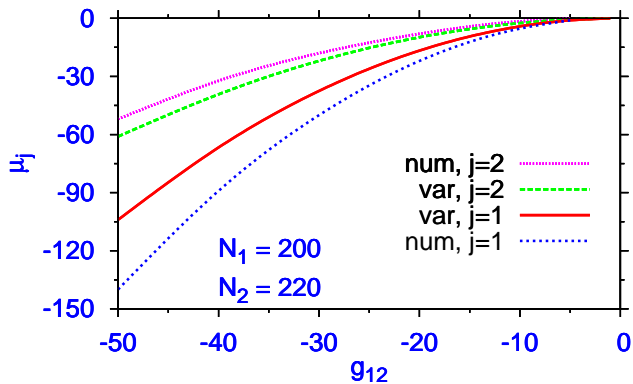}}
\end{center}
\caption{(Color online)
Chemical potential $\mu_j$ of $j$th component
obtained from the numerical solution 
(labelled ``num") of Eqs. 
(\ref{mu1}) and (\ref{mu2}) and   that obtained from the variational 
results (\ref{muG1}) and (\ref{muG2}) (labelled ``var") vs. interspecies 
coupling $g_{12}$
for $N_1=200, N_2=220$ and $\lambda=1$.  
}
\label{fig4}
\end{figure}

First we report results for stationary profiles of the localized 
Fermi-Fermi solitons formed in the presence of attractive interspecies 
interaction (negative $g_{12}$). The fermions 
form  BCS state(s) which satisfy a coupled 
nonlinear Schr\"odinger equation 
with repulsive (self defocusing) quintic nonlinearity. 
Hence fermions cannot form a bright soliton by itself. 
However, they can form a bright soliton in the presence of an
attractive interspecies  interaction \cite{skbs} induced by varying an 
external background magnetic field near a Feshbach resonance 
\cite{fesh}.

In Fig. \ref{fig3} we present the soliton profiles of the two 
components calculated by a direct numerical solution of Eqs. 
(\ref{mu1})  and (\ref{mu2}) and compare them with variational  results
(\ref{WG1}) and (\ref{WG2}). In 
general the numerical solutions have a profile distinct from a Gaussian 
shape of the variational approximation.
The numerical density profile reminds of a square barrier. 
Nevertheless, the variational approximation presents a faithful average 
description. From Figs.  \ref{fig3} (a) and (b) we find that for a fixed 
$N_1$ and $N_2$, as $|g_{12}|$ is increased, the solitons become more 
compact and are better represented by variational approximation. From 
Figs.  \ref{fig3} (b) and (c) we see that as the number of fermions is 
increased the numerical density profiles are more square-barrier type 
than a Gaussian type.  From Figs.  \ref{fig3} (c) and (d)
we find that for a fixed $g_{12}$, as the number of atoms is  reduced, 
the solitons become  more compact.

\begin{figure}[tbp]
\begin{center}
{\includegraphics[width=\linewidth]{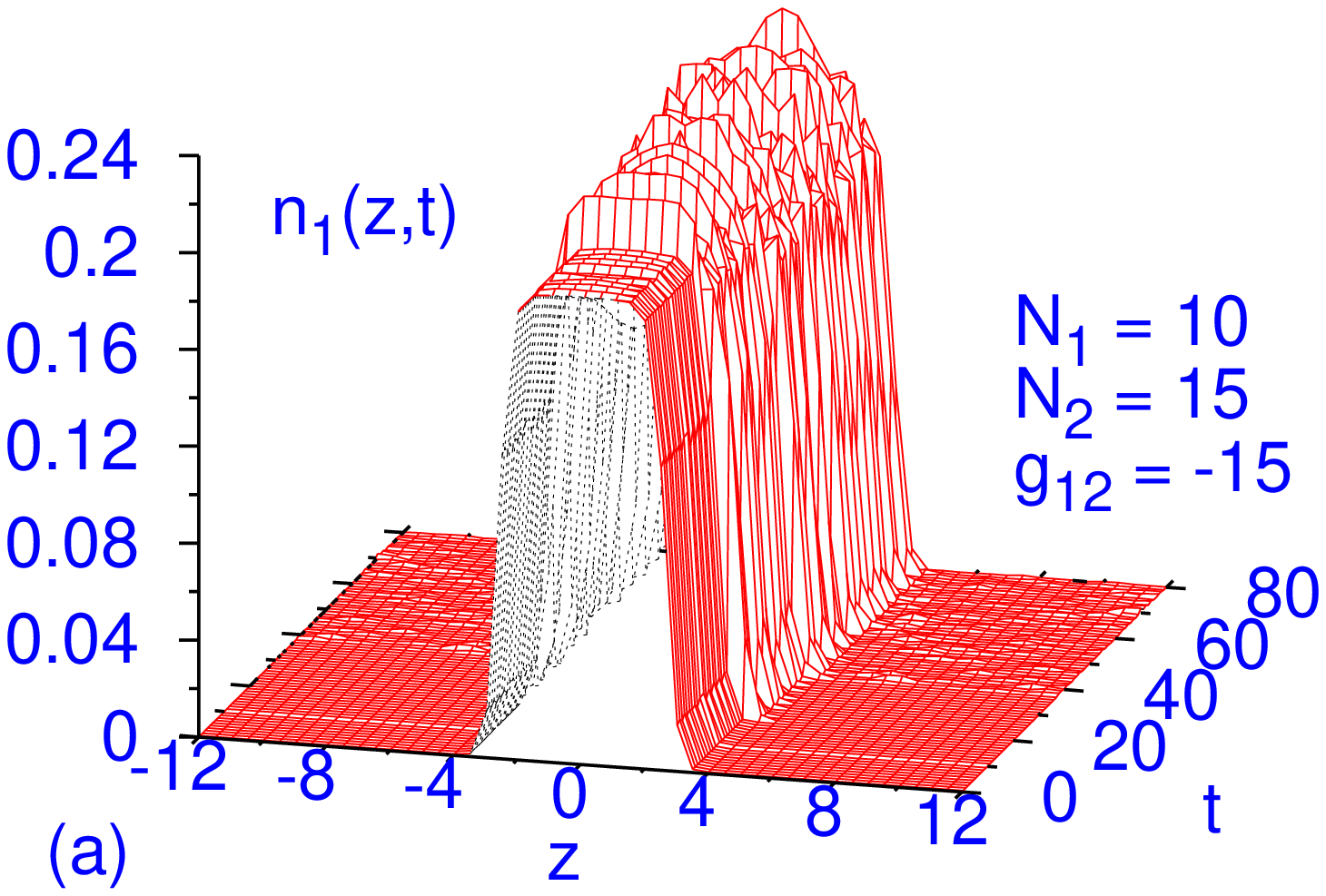}}
{\includegraphics[width=\linewidth]{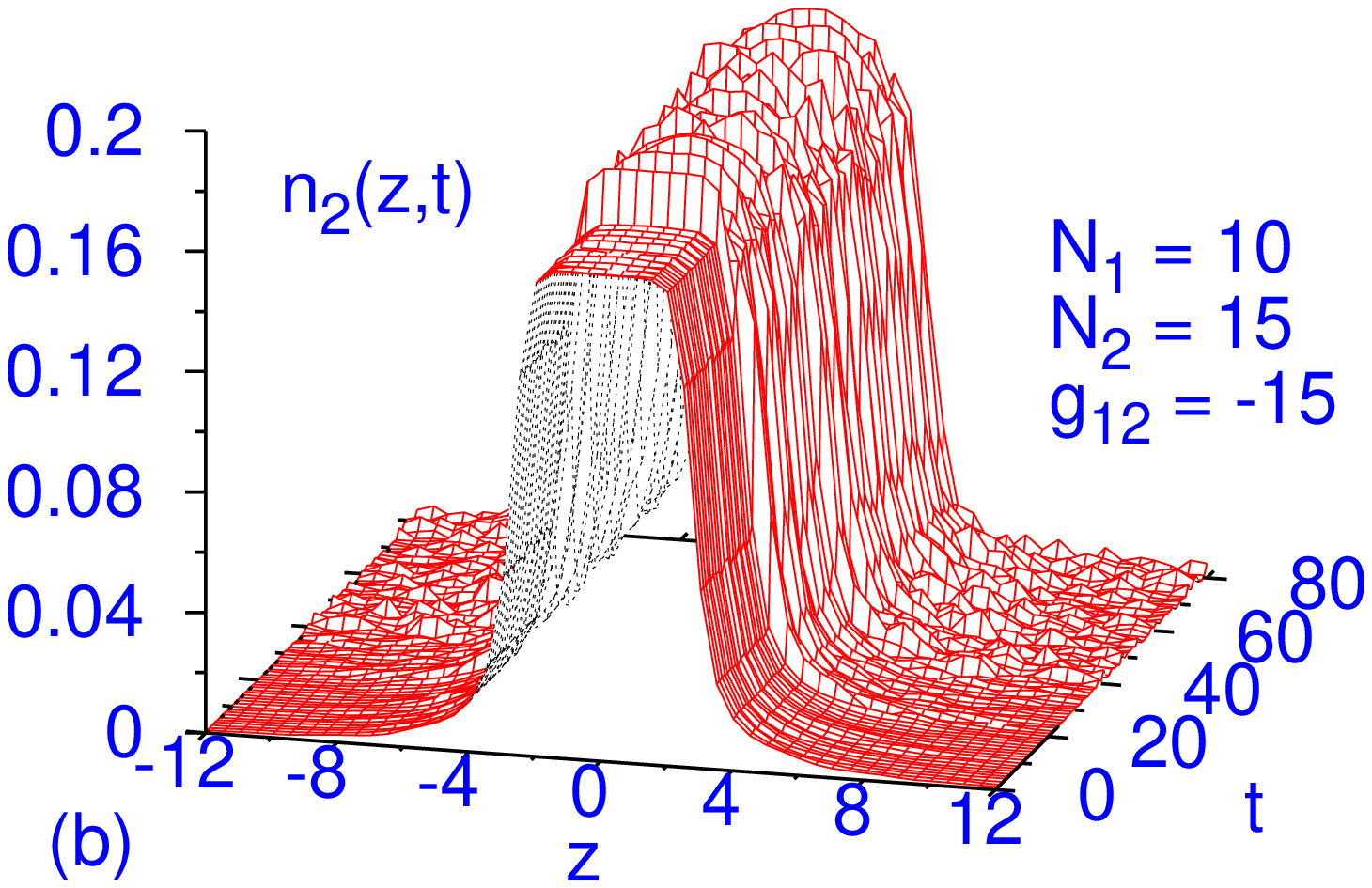}}
\end{center}
\caption{(Color online)
Dynamics of the probability density profiles of (a) the first and (b) 
the second Fermi
solitons of Fig
\ref{fig3} (a) when at $t=20$ they are subject to a 
perturbation by setting $\phi_j(z,t) = 1.05\times \phi_j(z,t).$
The solitons undergo stable propagation as long as we could continue
numerical simulation. The initial soliton profile is calculated with
imaginary time propagation algorithm and the dynamics studied with
real time propagation algorithm. The soliton profiles are normalized to
unity: $\int_{-\infty}^\infty n_j(z,t) dz =1$.
}\label{fig5}
\end{figure}

Next we illustrate how well are the variational 
approximations (\ref{muG1}) and (\ref{muG2}) for the chemical potential 
compared to the numerical results. 
In Fig. \ref{fig4} we plot the numerically 
obtained  chemical potential for $N_1=200,N_2=220,$ and $\lambda =1$ for 
different $g_{12}$ 
and compare with the variational result. We see that the overall  
agreement is good for all $g_{12}$, although it is better for small 
$|g_{12}|$.

After illustrating the soliton profiles in different states it is now 
pertinent to verify if these solitons are dynamically stable under 
perturbation. To this end we consider the typical stationary soliton of 
Fig. \ref{fig3} 
(a) (obtained by the imaginary time propagation method) and 
subject 
it to 
the  perturbation
 by setting $\phi_j(z,t) = 1.05 \times \phi_j(z,t)$  and 
observe 
the resultant dynamics (obtained by the real time propagation 
method).
The resultant dynamics  is illustrated in Fig. \ref{fig5}. 
The 
solitons under 
this perturbation execute some  oscillation, generate some noise, 
nevertheless   propagate for as 
long as the numerical simulation was continued without being 
destroyed. 
This demonstrates the stability of the solitons under perturbation. 
 {For very strong perturbation, as expected,  the 
solitons are 
destroyed. If time covered by numerical simulation is too short, a 
unstable
solution might appear to be stable. Thus, in  numerical simulation
it is important to cover times large compared to characteristic
timescale of the problem, as in Fig. \ref{fig5}. Also,  a false 
stability might appear for a small interval of time for specific space 
and time steps used in discretization. We checked the stability for 
different time steps over large intervals of time.}

\section{Summary}

In this paper we have obtained  the phase diagram of a BCS superfluid 
Fermi-Fermi 
mixture of distinct mass fermions
at zero temperature in 1D, 2D, and 3D. The linear stability 
conditions relating the strength of interspecies Fermi-Fermi 
interaction  with the two Fermi densities  are 
obtained from an energetic consideration. Two possible equilibrium 
scenarios emerge: a uniform mixture 
and two pure separated phases. 
In 1D, two pure and separated phases appear for small fermion densities; 
for large densities appears the uniform mixture from an energetic 
consideration as shown in Fig. \ref{fig1}. 
In 3D, the opposite happens.  In addition, in 3D, a mixed and a pure
phase can appear.
In 2D, the conditions for uniform mixture 
and separated phases do not put any restriction on the  fermion 
densities but only on the interspecies Fermi-Fermi interaction. 

In 3D, the uniform mixture is unstable against small fluctuations for
large Fermi densities  for a fixed $g_{12}^2$. 
For a positive $g_{12}$ it
should show partial demixing and for a negative $g_{12}$ it may undergo 
collapse. 
In 1D, the uniform mixture is unstable against small fluctuations for 
small Fermi densities for a fixed $g_{12}^2$. For a positive $g_{12}$ it 
should show partial demixing and for a negative $g_{12}$ it should form 
bright 
solitons. Hence  
this mixture is of special interest for a negative $g_{12}$.
This is the domain of soliton formation by modulational instability of 
the  uniform mixture. 
To study the modulational instability and soliton formation in the 
mixture we derive a set of coupled nonlinear equations  
derived as the Euler-Lagrange 
equation employing the Lagrangian density of the mixture. The condition 
of modulational instability so obtained is consistent with that of  
stability of uniform mixture obtained from an energetic consideration.
In addition, 
we solve the  1D dynamical equations numerically and variationally to 
study the density and chemical potential of the solitons. The 
variational result is found to be in good agreement with the numerical 
solution. 
We also 
established numerically the dynamical stability of the Fermi-Fermi 
solitons by subjecting them to  a perturbation  by 
multiplying the 
wave-function profiles by 1.05. The system is then found to propagate 
over a very long period of time  without 
being destroyed, which demonstrated the stability of the solitons. 

\acknowledgments

We  thank  Dr. Luca Salasnich for comments and discussion and 
FAPESP and CNPq for partial financial support.

\end{document}